\documentclass[11pt,a4paper,twosided]{article}
\usepackage{geometry}
\usepackage[utf8]{inputenc}
\usepackage[T1]{fontenc}

\usepackage[english]{babel}

\usepackage{amsmath,amsthm,amssymb,stmaryrd,thmtools,upgreek}
\newtheorem{theorem}{Theorem}
\newtheorem{lemma}[theorem]{Lemma}
\newtheorem{corollary}[theorem]{Corollary}

\theoremstyle{definition}
\newtheorem{definition}[theorem]{Definition}
\newtheorem{remark}[theorem]{Remark}
\newtheorem{example}[theorem]{Example}

\usepackage{graphicx}
\usepackage[obeyclassoptions,mode=tex]{standalone}
\usepackage{tikz}
\usetikzlibrary{backgrounds}
\usetikzlibrary{shapes.geometric}
\usetikzlibrary{positioning}
\usetikzlibrary{automata}
\usetikzlibrary{tikzmark}
\usetikzlibrary{patterns}
\usetikzlibrary{arrows}
\tikzset{every state/.style={minimum size=1pt}}
\usepackage{tikz-cd}

\usepackage{hyperref}
\usepackage[capitalise,noabbrev,nameinlink]{cleveref}
\usepackage[electronic,hyperref,xcolor,cleveref]{knowledge}
\knowledgeconfigure{notion}

\usepackage{booktabs}
\usepackage{varwidth}

\usepackage{xparse}
\usepackage{xpatch}
\usepackage{tokcycle}
\usepackage{ifthen}

\usepackage{subcaption} %

\usepackage{ensps-colorscheme}

\hypersetup{
    colorlinks=true,
    linkcolor=Prune,
    filecolor=Prune,
    urlcolor=Prune,
    citecolor=Prune,
}
\knowledgestyle{intro notion}{color=B4,emphasize}
\knowledgestyle{notion}{color=A4}

\makeatletter
\newcommand\mathgr[1]{\tokcycle
  {\addcytoks{##1}}
  {\processtoks{##1}}
  {\ifcsname up\expandafter\@gobble\string##1\endcsname
   \addcytoks[1]{\csname up\expandafter\@gobble\string##1\endcsname}%
    \else\addcytoks{##1}\fi}
  {\addcytoks{##1}}{#1}%
  \expandafter\mathrm\expandafter{\the\cytoks}%
}
\makeatother

\NewDocumentEnvironment{proofof}{ m O{appendix} }{
    \ifcsname #1\endcsname
        \def\isInsideRestatedTheorem{1}
        \csname #1\endcsname*
    \fi
    \begin{proof}[Proof of {\cref{#1}} as stated on page {\pageref{#1}}]
        \phantomsection
        \label{#1:proof}
}{
        \ifthenelse{\equal{#2}{appendix}}{
        \marginpar{\vspace{-2em}\texttt{\small{\hyperref[#1]{$\triangleright$ Back to p.\pageref{#1}}}}}
        }{}
    \end{proof}
}

\NewDocumentCommand{\proofref}{ m }{
    \IfRefUndefinedExpandable{#1:proof}{}{
        \ifdefined\isInsideRestatedTheorem
        \else
            \marginpar{\vspace{0.6em}\texttt{\small{\hyperref[#1:proof]{$\triangleright$ Proven p.\pageref{#1:proof}}}}}
        \fi
    }
}

\NewDocumentCommand{\NewDocumentOrdering}{ m m m }{
    \expandafter\newcommand\csname #1leq\endcsname{
        \mathrel{\kl[#1]{#2}}
    }
    \expandafter\newcommand\csname #1lt\endcsname{
        \mathrel{\kl[#1]{#3}}
    }
    \knowledge{#1}{notion}
}

\NewDocumentCommand{\set}{ m }{\{ #1 \}}
\NewDocumentCommand{\setof}{ m m }{\{ #1 \mid #2 \}}
\NewDocumentCommand{\card}{ m }{\left| #1 \right|}
\NewDocumentCommand{\Nat}{ }{\mathbb{N}}
\NewDocumentCommand{\Rel}{ }{\mathbb{Z}}
\NewDocumentCommand{\seqof}{ m O{n \in \Nat} }{\left( #1 \right)_{#2}}
\NewDocumentCommand{\defined}{ }{\triangleq}

\NewDocumentCommand{\range}{ O{1} m }{[#1, #2]}

\NewDocumentCommand{\upset}{ O{} m }{{\uparrow_{#1} #2}}
\NewDocumentCommand{\dwset}{ O{} m }{{\downarrow_{#1} #2}}

\NewDocumentCommand{\factorial}{ O{} m }{
    \if\relax\detokenize{#1}\relax
        #2!
    \else
        (#2)!
    \fi
}
\NewDocumentOrdering{pref}{\sqsubseteq_{\mathsf{pref}}}{\sqsubset_{\mathsf{pref}}}
\NewDocumentOrdering{suff}{\sqsubseteq_{\mathsf{suff}}}{\sqsubset_{\mathsf{suff}}}
\NewDocumentOrdering{inf}{\sqsubseteq_{\mathsf{infix}}}{\sqsubset_{\mathsf{infix}}}
\NewDocumentOrdering{hig}{\leq^*}{<^*}
\NewDocumentOrdering{run}{\trianglelefteq}{\trianglelefts}

\NewDocumentCommand{\InfPeriodChain}{ m }{
    \mathop{\kl[\InfPeriodChain]{\mathsf{PI}\!\!\!\!\downarrow}}
    ( #1 )
}
\knowledge{\InfPeriodChain}{notion}

\NewDocumentCommand{\canrun}{O{can}}{
    \mathop{\mathsf{#1}}
}
\NewDocumentCommand{\yieldrun}{}{
    \mathop{\kl[\yieldrun]{\mathsf{yield}}}
}
\knowledge{\yieldrun}{notion}

\NewDocumentCommand{\HigEmb}{}{
    \mathop{\kl[\HigEmb]{\mathsf{Hom}^*}}
}
\knowledge{\HigEmb}{notion}

\NewDocumentCommand{\GapWord}{ m m }{
    \mathop{\kl[\GapWord]{\mathsf{G}}}_{#2}^{#1}
}
\knowledge{\GapWord}{notion}

\NewDocumentCommand{\GapLanguage}{ m m }{
    \mathop{\kl[\GapLanguage]{\mathsf{L}}}_{#2}^{#1}
}
\knowledge{\GapLanguage}{notion}

\NewDocumentCommand{\ThueMorse}{}{\kl[\ThueMorse]{\mathbf{t}}}
\knowledge{\ThueMorse}{notion}

\NewDocumentCommand{\LMorse}{}{\kl[\LMorse]{I_\mathbf{t}}}
\knowledge{\LMorse}{notion}

\NewDocumentCommand{\MSO}{ }{\kl[\MSO]{\mathsf{MSO}}}
\knowledge{\MSO}{notion}

\NewDocumentCommand{\cansep}{}{\#}
\NewDocumentCommand{\project}{}{\pi}

\NewDocumentCommand{\oType}{ m }{\kl[\oType]{\mathfrak{o}}(#1)}
\knowledge{\oType}{notion}
\NewDocumentCommand{\oHeight}{ m }{\kl[\oHeight]{\mathfrak{h}}(#1)}
\knowledge{\oHeight}{notion}
\NewDocumentCommand{\oWidth}{ m }{\kl[\oWidth]{\mathfrak{w}}(#1)}
\knowledge{\oWidth}{notion}

\NewDocumentCommand{\oComProd}{}{\mathop{\kl[\oComProd]{\otimes}}}
\knowledge{\oComProd}{notion}

\NewDocumentCommand{\omegaOrd}{ }{\kl[\omegaOrd]{\omega}}
\knowledge{\omegaOrd}{notion}

\NewDocumentCommand{\omegaOne}{ }{\kl[\omegaOne]{\omega_1}}
\knowledge{\omegaOne}{notion}

\NewDocumentCommand{\infset}{ m }{\kl[\infset]{\mathsf{Infixes}}(#1)}
\knowledge{\infset}{notion}
\knowledge{url=https://autoboz.org/}
 | Autobóz

\knowledge{notion}
 | quasi-order
\knowledge{notion}
 | ordinal
\knowledge{notion}
 | totally ordered
 | total order
\knowledge{notion}
 | well-quasi-ordered
 | well-quasi-ordering
 | well-quasi-orderings
 | wqo
 | wqos
 | well-quasi-order
 | well-quasi-orders
\knowledge{notion}
 | partial ordering
 | partial order
 | partially ordered
\knowledge{notion}
 | well-founded
\knowledge{notion}
 | good@sequence
 | good@wqo
 | good sequence
\knowledge{notion}
 | bad@sequence
 | bad sequences
 | bad sequence
\knowledge{notion}
 | downwards closure
 | downwards closures
\knowledge{notion}
 | downwards closed
 | Downwards Closed
\knowledge{notion}
 | upwards closure
\knowledge{notion}
 | decreasing sequence
 | decreasing sequences
\knowledge{notion}
 | antichain
 | antichains
\knowledge{notion}
 | chains
 | chain
\knowledge{notion}
 | order embedding
 | embeds into
 | embedding@order
 | embedding
 | embeddings
 | embedded
\knowledge{notion}
 | subword relation
 | scattered subword
 | scattered subword relation
 | subword
 | subwords
 | Higman’s subword relation
 | subword embedding
 | subword embedding relation
 | subword embeddings

\knowledge{notion}
 | infix relation
 | infix ordering
 | infix
 | infixes 
\knowledge{notion}
 | prefix relation
 | prefix ordering
 | prefix 
 | prefixes
\knowledge{notion}
 | suffix relation
 | suffix ordering 
 | suffix 
 | suffixes
\knowledge{notion}
 | ordinal invariants
\knowledge{notion}
 | ordinal height
\knowledge{notion}
 | maximal order type
 | m.o.t.
\knowledge{notion}
 | ordinal width
\knowledge{notion}
 | periodic word
 | periodic words
 | periodic@word
\knowledge{notion}
 | periodic length
 | periodic lengths

\knowledge{notion}
 | regular language
 | regular@language
 | regular languages
\knowledge{notion}
 | finite automaton
 | finite automata
\knowledge{notion}
 | rational transduction
 | rational transductions
\knowledge{notion}
 | closed under rational transductions
 | effectively closed under rational transductions

\knowledge{notion}
 | $\MSO $-definable
\knowledge{notion}
 | $\MSO $
 | $\MSO $-formula
\knowledge{notion}
 | Thue-Morse sequence
\knowledge{notion}
 | cube-free
\knowledge{notion}
 | uniformly recurrent
 | uniformly recurrent words

\knowledge{notion}
 | ultimately uniformly recurrent

\knowledge{notion}
 | antichain branch
 | antichain branches

\knowledge{notion}
 | amalgamation system
 | amalgamation systems

\knowledge{notion}
 | effective amalgamation system
 | effective amalgamation systems

\knowledge{notion}
 | amalgamation language
 | language@amalgamation

\knowledge{notion}
 | tree of prefixes

\knowledge{notion}
 | gap word
 | gap words
\knowledge{notion}
 | gap language
 | gap languages
\knowledge{notion}
 | unbounded@language
 | unbounded language
 | unbounded languages
\knowledge{notion}
 | bounded language
 | bounded@language
 | bounded languages

\knowledge{notion}
 | effective@amalg
 | effective amalgamative class
 | effective amalgamative classes

\knowledge{notion}
 | strongly effective@amalg
 | strongly effective amalgamative class
 | strongly effective amalgamative classes

\knowledge{notion}
 | order ideal
 | order ideals
 | ideal@order
 | ideals@order

\knowledge{notion}
 | directed subsets
 | directed subset
 | directed@subset

\knowledge{notion}
 | Automatic@sequence
 | automatic sequences
 | automatic@sequence
 | automatic sequence

\knowledge{notion}
 | canonical decomposition
\knowledge{notion}
 | admissible embeddings

\knowledge{notion}
 | commutative ordinal product
 | Hessenberg product

\title{Well-quasi-orderings on word languages}
\author{%
        Nathan Lhote\thanks{Aix-Marseille University}
     \and
        Aliaume Lopez\thanks{University of Warsaw}
     \and
        Lia Schütze\thanks{MPI-SWS, Germany}
    }

\newcommand{\makeabstract}{
\begin{abstract}
    The set of finite words over a well-quasi-ordered set is itself
    well-quasi-ordered. This seminal result by Higman is a cornerstone
    of the theory of well-quasi-orderings and has found numerous
    applications in computer science. However, this result is based on a
    specific choice of ordering on words, the (scattered) subword
    ordering. In this paper, we describe to what extent other natural
    orderings (prefix, suffix, and infix) on words can be used to derive
    Higman-like theorems. More specifically, we are interested in
    characterizing \emph{languages} of words that are well-quasi-ordered
    under these orderings. We show that a simple characterization is
    possible for the prefix and suffix orderings, and that under extra
    regularity assumptions, this also extends to the infix ordering.
\end{abstract}
}

\begin{document}
\maketitle
\makeabstract

\section{Introduction}
\label{introduction:sec}

\AP A \intro{well-quasi-ordered} set is a set $X$ equipped with a quasi-order
$\preceq$ such that every infinite sequence $\seqof{x_n}$ of elements taken in
$X$ contains an increasing pair $x_i \preceq x_j$ with $i < j$. Well-quasi-orderings serve
as a core combinatorial tool powering many termination arguments, and was
successfully applied to the verification of infinite state transition
systems~\cite{ABDU96,ABDU98}. One of the appealing properties of
well-quasi-orderings is that they are closed under many operations, such as
taking products, finite unions, and finite powerset
constructions~\cite{SCSC12}. Perhaps more surprisingly, the class of
well-quasi-ordered sets is also stable under the operation of taking finite
words and finite trees labelled by elements of a well-quasi-ordered set
\cite{HIG52,KRU72}.

\AP Note that in the case of finite words and finite trees, the precise choice
of ordering is crucial to ensure that the resulting structure is
\kl{well-quasi-ordered}. The celebrated result of Higman states that the set of
finite words over an ordered alphabet $(X, \preceq)$ is \kl{well-quasi-ordered}
by the so-called \kl{subword embedding relation}~\cite{HIG52}. Let us recall
that the \kl{subword relation} for words over $(X, \preceq)$ is defined as
follows: a word $u$ is a \intro{subword} of a word $v$, written $u
\intro*\higleq v$, if there exists an increasing function $f \colon \{1,
\ldots, |u|\} \to \{1, \ldots, |v|\}$ such that $u_i \preceq v_{f(i)}$ for all
$i \in \{1, \ldots, |u|\}$.

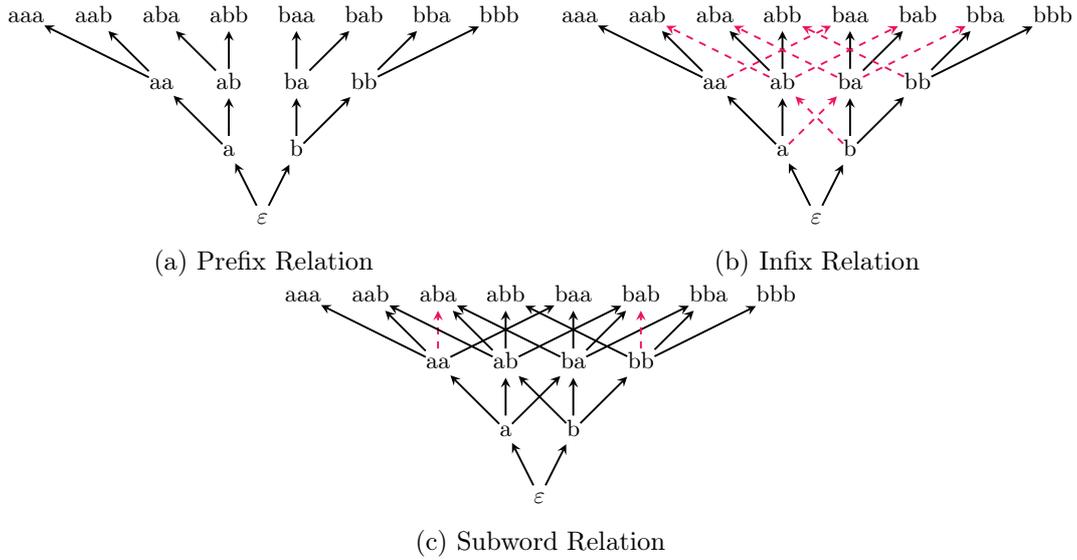
\begin{figure}
    \centering
    \begin{subfigure}[t]{0.48\textwidth}
    	\centering
    	\begin{tikzpicture}[
        isWord/.style={rectangle,inner sep=0mm},
        isEmbedding/.style={->, >=stealth, thick},
        branch/.style={A5},
        antichain/.style={A2},
        branchEdge/.style={A5,dashed},
        outEdge/.style={dotted,B2},
        isInfixEnc/.style={},
        isRealWord/.style={A4,opacity=0.2},
        isRealEmbedding/.style={A4,opacity=0.2},
        isNewEdge/.style={B2,dashed},
        ]

\node[isWord] (epsilon) at (-0.5, 0) {\strut $\varepsilon$};
\node[isWord] (a) at (-1.0, 1) {\strut a};
\node[isWord] (b) at (0.0, 1) {\strut b};
\node[isWord] (aa) at (-2.0, 2) {\strut aa};
\node[isWord] (ab) at (-1.0, 2) {\strut ab};
\node[isWord] (ba) at (0.0, 2) {\strut ba};
\node[isWord] (bb) at (1.0, 2) {\strut bb};
\node[isWord] (aaa) at (-4.0, 3) {\strut aaa};
\node[isWord] (aab) at (-3.0, 3) {\strut aab};
\node[isWord] (aba) at (-2.0, 3) {\strut aba};
\node[isWord] (abb) at (-1.0, 3) {\strut abb};
\node[isWord] (baa) at (0.0, 3) {\strut baa};
\node[isWord] (bab) at (1.0, 3) {\strut bab};
\node[isWord] (bba) at (2.0, 3) {\strut bba};
\node[isWord] (bbb) at (3.0, 3) {\strut bbb};
\draw[isEmbedding] (epsilon) -- (a);
\draw[isEmbedding] (epsilon) -- (b);
\draw[isEmbedding] (a) -- (aa);
\draw[isEmbedding] (a) -- (ab);
\draw[isEmbedding] (b) -- (ba);
\draw[isEmbedding] (b) -- (bb);
\draw[isEmbedding] (aa) -- (aaa);
\draw[isEmbedding] (aa) -- (aab);
\draw[isEmbedding] (ab) -- (aba);
\draw[isEmbedding] (ab) -- (abb);
\draw[isEmbedding] (ba) -- (baa);
\draw[isEmbedding] (ba) -- (bab);
\draw[isEmbedding] (bb) -- (bba);
\draw[isEmbedding] (bb) -- (bbb);

\end{tikzpicture}
    	\caption{Prefix Relation}
   	\end{subfigure}%
   	\hfill%
   	\begin{subfigure}[t]{0.48\textwidth}
   		\centering
   		\begin{tikzpicture}[
        isWord/.style={rectangle,inner sep=0mm},
        isEmbedding/.style={->, >=stealth, thick},
        branch/.style={A5},
        antichain/.style={A2},
        branchEdge/.style={A5,dashed},
        outEdge/.style={dotted,B2},
        isInfixEnc/.style={},
        isRealWord/.style={A4,opacity=0.2},
        isRealEmbedding/.style={A4,opacity=0.2},
        isNewEdge/.style={B2,dashed},
        ]

\node[isWord] (epsilon) at (-0.5, 0) {\strut $\varepsilon$};
\node[isWord] (a) at (-1.0, 1) {\strut a};
\node[isWord] (b) at (0.0, 1) {\strut b};
\node[isWord] (aa) at (-2.0, 2) {\strut aa};
\node[isWord] (ab) at (-1.0, 2) {\strut ab};
\node[isWord] (ba) at (0.0, 2) {\strut ba};
\node[isWord] (bb) at (1.0, 2) {\strut bb};
\node[isWord] (aaa) at (-4.0, 3) {\strut aaa};
\node[isWord] (aab) at (-3.0, 3) {\strut aab};
\node[isWord] (aba) at (-2.0, 3) {\strut aba};
\node[isWord] (abb) at (-1.0, 3) {\strut abb};
\node[isWord] (baa) at (0.0, 3) {\strut baa};
\node[isWord] (bab) at (1.0, 3) {\strut bab};
\node[isWord] (bba) at (2.0, 3) {\strut bba};
\node[isWord] (bbb) at (3.0, 3) {\strut bbb};
\draw[isEmbedding] (epsilon) -- (a);
\draw[isEmbedding] (epsilon) -- (b);
\draw[isEmbedding] (a) -- (aa);
\draw[isEmbedding] (a) -- (ab);
\draw[isEmbedding,isNewEdge] (a) -- (ba);
\draw[isEmbedding,isNewEdge] (b) -- (ab);
\draw[isEmbedding] (b) -- (ba);
\draw[isEmbedding] (b) -- (bb);
\draw[isEmbedding] (aa) -- (aaa);
\draw[isEmbedding] (aa) -- (aab);
\draw[isEmbedding,isNewEdge] (aa) -- (baa);
\draw[isEmbedding,isNewEdge] (ab) -- (aab);
\draw[isEmbedding] (ab) -- (aba);
\draw[isEmbedding] (ab) -- (abb);
\draw[isEmbedding,isNewEdge] (ab) -- (bab);
\draw[isEmbedding,isNewEdge] (ba) -- (aba);
\draw[isEmbedding] (ba) -- (baa);
\draw[isEmbedding] (ba) -- (bab);
\draw[isEmbedding,isNewEdge] (ba) -- (bba);
\draw[isEmbedding,isNewEdge] (bb) -- (abb);
\draw[isEmbedding] (bb) -- (bba);
\draw[isEmbedding] (bb) -- (bbb);

\end{tikzpicture}
   		\caption{Infix Relation}
   	\end{subfigure}
   	\begin{subfigure}[t]{0.48\textwidth}
   		\centering
   		\begin{tikzpicture}[
        isWord/.style={rectangle,inner sep=0mm},
        isEmbedding/.style={->, >=stealth, thick},
        branch/.style={A5},
        antichain/.style={A2},
        branchEdge/.style={A5,dashed},
        outEdge/.style={dotted,B2},
        isInfixEnc/.style={},
        isRealWord/.style={A4,opacity=0.2},
        isRealEmbedding/.style={A4,opacity=0.2},
        isNewEdge/.style={B2,dashed},
        ]

\node[isWord] (epsilon) at (-0.5, 0) {\strut $\varepsilon$};
\node[isWord] (a) at (-1.0, 1) {\strut a};
\node[isWord] (b) at (0.0, 1) {\strut b};
\node[isWord] (aa) at (-2.0, 2) {\strut aa};
\node[isWord] (ab) at (-1.0, 2) {\strut ab};
\node[isWord] (ba) at (0.0, 2) {\strut ba};
\node[isWord] (bb) at (1.0, 2) {\strut bb};
\node[isWord] (aaa) at (-4.0, 3) {\strut aaa};
\node[isWord] (aab) at (-3.0, 3) {\strut aab};
\node[isWord] (aba) at (-2.0, 3) {\strut aba};
\node[isWord] (abb) at (-1.0, 3) {\strut abb};
\node[isWord] (baa) at (0.0, 3) {\strut baa};
\node[isWord] (bab) at (1.0, 3) {\strut bab};
\node[isWord] (bba) at (2.0, 3) {\strut bba};
\node[isWord] (bbb) at (3.0, 3) {\strut bbb};
\draw[isEmbedding] (epsilon) -- (a);
\draw[isEmbedding] (epsilon) -- (b);
\draw[isEmbedding] (a) -- (aa);
\draw[isEmbedding] (a) -- (ab);
\draw[isEmbedding] (a) -- (ba);
\draw[isEmbedding] (b) -- (ab);
\draw[isEmbedding] (b) -- (ba);
\draw[isEmbedding] (b) -- (bb);
\draw[isEmbedding] (aa) -- (aaa);
\draw[isEmbedding] (aa) -- (aab);
\draw[isEmbedding,isNewEdge] (aa) -- (aba);
\draw[isEmbedding] (aa) -- (baa);
\draw[isEmbedding] (ab) -- (aab);
\draw[isEmbedding] (ab) -- (aba);
\draw[isEmbedding] (ab) -- (abb);
\draw[isEmbedding] (ab) -- (bab);
\draw[isEmbedding] (ba) -- (aba);
\draw[isEmbedding] (ba) -- (baa);
\draw[isEmbedding] (ba) -- (bab);
\draw[isEmbedding] (ba) -- (bba);
\draw[isEmbedding] (bb) -- (abb);
\draw[isEmbedding,isNewEdge] (bb) -- (bab);
\draw[isEmbedding] (bb) -- (bba);
\draw[isEmbedding] (bb) -- (bbb);

\end{tikzpicture}
   		\caption{Subword Relation}
   	\end{subfigure}
   	
   	\caption{A simple representation of the \kl{subword relation},
        \kl{prefix relation},
        and \kl{infix relation},
        on the alphabet $\set{a,b}$ for words of
        length at most $3$. The figures are Hasse Diagrams,
        representing the successor relation of the order.
        Furthermore, we highlight in dashed red relations that are added
        when moving from the prefix relation to the infix one,
        and to the infix relation to the subword one.}
    \label{word-embeddings:fig}
\end{figure}

\AP However, there are many other natural orderings on words that could be
considered in the context of \kl{well-quasi-orderings}, even in the simplified
setting of a finite alphabet $\Sigma$ equipped with the equality relation. In
this setting, the three alternatives we consider are the \intro{prefix
relation} ($u \intro*\prefleq v$ if there exists $w$ with $uw = v$), the
\intro{suffix relation} ($u \intro*\suffleq v$ if there exists $w$ such that
$wu = v$), and the \intro{infix relation} ($u \intro*\infleq v$ if there exists
$w_1,w_2$ such that $w_1 u w_2 = v$). Note that these three relations
straightforwardly generalize to infinite quasi-ordered alphabets.
Unfortunately, it is easy to see that none of these constructions are 
well-quasi-ordered as soon as the alphabet contains two distinct letters:
for instance, the infinite sequence $a^n b^n a^n$ is \kl{well-quasi-ordered} by
the \kl{subword relation} but by neither the \kl{prefix relation}, nor the
\kl{suffix relation}, nor the \kl{infix relation}.

\AP While this dooms \kl[well-quasi-ordered]{well-quasi-orderedness} of these
relations in the general case, there may be \emph{subsets} of $\Sigma^*$ which
are \kl{well-quasi-ordered} by these relations. As a simple example, take the
case of finite sets of (finite) words which are all \kl{well-quasi-ordered}
regardless of the ordering considered. This raises the question of
characterizing exactly which subsets $L \subseteq \Sigma^*$ are
\kl{well-quasi-ordered} with respect to the \kl{prefix relation} (respectively,
the \kl{suffix relation} or the \kl{infix relation}), and designing
suitable decision procedures.

\AP Let us argue that these decision procedures fit a larger picture in the
research area of well-quasi-orderings.
Indeed, there has been recent breakthroughs in deciding whether a given order
is a \kl{well-quasi-order}, for instance in the context of the verification of
infinite state transition systems~\cite{DBLP:conf/fsttcs/FinkelG19} or in the
context of logic~\cite{DBLP:journals/pacmpl/BergstrasserGLZ24}.
Furthermore, a previous work by Kuske shows that any
\emph{reasonable}\footnote{ This will be made precise in
\cref{infix-embedding:thm}. } partially ordered set $(X, \leq)$ can
be embedded into $\set{a,b}^*$ with the \kl{infix relation}~\cite[Lemma
5.1]{DBLP:journals/ita/Kuske06}. Phrased differently, one can encode a large
class of partially ordered sets as subsets of $\set{a,b}^*$. As a consequence,
the following decision problem provides a reasonable abstract framework for
deciding whether a given partially ordered set is \kl{well-quasi-ordered}:
given a language $L \subseteq \Sigma^*$, decide whether $L$ is
\kl{well-quasi-ordered} by the \kl{infix relation}.

\AP When considering an algorithm based on a \kl{well-quasi-ordering}, the
runtime of the algorithm is deeply related to the ``complexity'' of the
underlying quasi-order~\cite{SCHMITZ17}. One way to measure this complexity is
to consider its so-called \kl{ordinal invariants}: for instance, the
\kl{maximal order type} (or \kl{m.o.t.}), originally defined by De Jongh and Parikh
\cite{dejongh77}, is the order type of the maximal linearization of a
well-quasi-ordered set. In the case of a finite set, the \kl{m.o.t.} is precisely
the size of the set. Better runtime bounds were obtained by considering two
other parameters~\cite{SCHMITZ19}: the \kl{ordinal height} introduced by
Schmidt \cite{schmidt81}, and the \kl{ordinal width} of Kříž and Thomas
\cite{kriz90b}. Therefore, when characterizing \kl{well-quasi-ordered}
languages, we will also be interested in deriving upper bounds on their
\kl{ordinal invariants}. We refer to \cref{ordinal-invariants:subsec} for a
more detailed introduction to these parameters and ordinal computations in
general.

\subparagraph{Contribution} In this paper, we focus on words over a finite
alphabet $\Sigma$, and characterize subsets $L \subseteq \Sigma^*$ that are
\kl{well-quasi-ordered} by the \kl{prefix relation}, the \kl{suffix relation},
and the \kl{infix relation}. Furthermore, we devise decision algorithms
whenever the languages are given by reasonable computational models. Finally,
we provide precise bounds on the possible \kl{ordinal invariants} of such
languages.

In the case of the \kl{prefix} and \kl{suffix} relations, we show that a
language $L \subseteq \Sigma^*$ is \kl{well-quasi-ordered} by the \kl{prefix
relation} (resp. \kl{suffix}) if and only if it is a finite union of
\kl{chains} of the \kl{prefix relation} (resp. \kl{prefix}), where a
\intro{chain} is a totally ordered set that is well-founded. Note that
\kl{chains} are the simplest possible \kl{well-quasi-ordered} sets (they are
\kl{totally ordered} and \kl{well-founded}) and it is known that finite unions
of \kl{well-quasi-ordered} sets are \kl{well-quasi-ordered}. As a consequence,
the above characterization states that only the simplest possible
\kl{well-quasi-ordered} sets are \kl{well-quasi-ordered} by the \kl{prefix
relation} (\cref{prefixes:thm}). Let us furthermore highlight that this
characterization holds without any restriction on the decidability of the
language $L$ itself, but heavily relies on the assumption that $\Sigma$ is
finite. This allows us to derive tight bounds on the \kl{ordinal invariants} of
such \kl{well-quasi-ordered} languages, which can be interpreted in two dual
ways: first, these languages are extremely simple, which means that one could
have hoped for using directly another well-quasi-ordered set without resorting
to finite words; second, any time one encounters a \kl{well-quasi-ordered}
language, it is going to provide relatively efficient algorithms. Indeed, we
prove that all languages $L \subseteq \Sigma^*$ are \kl{well-quasi-ordered} by
the \kl{prefix relation} (resp. \kl{suffix relation}) have an \kl{ordinal
height} of at most $\omegaOrd$, a finite \kl{ordinal width}, and a \kl{maximal
order type} strictly below $\omegaOrd^2$
(\cref{prefixes-ordinal-invariants:cor}).

\AP The straightforward generalization of the results for the \kl{prefix
relation} and the \kl{suffix relation} to the \kl{infix relation} is not
possible. Indeed, it follows from Kuske's result that $\Sigma^*$ equipped with
the \kl{subword relation} can be embedded into $\set{a,b}^*$ with the \kl{infix
relation}~\cite[Lemma 5.1]{DBLP:journals/ita/Kuske06}. This implies that there
are \kl{well-quasi-ordered} languages for the \kl{infix relation} that have
very large \kl{ordinal invariants}: for instance, the \kl{maximal order type}
of the \kl{subword relation} is $\omegaOrd^{\omegaOrd^{\card{\Sigma} - 1}}$,
which equals its \kl{ordinal width} \cite[Corollary 3.9, Theorem
4.21]{DZSCSC20}. We show that in two situations, this can be avoided: when the
language is \kl{downwards closed} (i.e., when it is closed under taking
\kl{infixes}), and when the language is \kl(language){bounded} (i.e., 
when it is included in some $w_1^* \cdots w_k^*$ for some finite choice of
words $w_1, \ldots, w_k$). In those cases, we are able to characterize
\kl{well-quasi-ordered} languages by the \kl{infix relation} and derive tight
bounds on their \kl{ordinal invariants}.

In the case of \kl{bounded languages}, we prove that a \kl{bounded language} $L
\subseteq \Sigma^*$ is \kl{well-quasi-ordered} by the \kl{infix relation} if
and only if it is (included in) a finite union of languages $S_i \cdot P_i$,
where each $S_i$ is a \kl{chain} for the \kl{suffix relation}, and where each
$P_i$ is a \kl{chain} for the \kl{prefix relation}
(\cref{bounded-language:thm}) This result directly translates into upper bounds
on the possible \kl{ordinal invariants} of such languages similarly as for the
\kl{prefix relation}. Notice that these upper bounds are significantly smaller
than \kl{ordinal invariants} of the \kl{subword relation}: they have an \kl{ordinal
height} of at most $\omegaOrd$, an \kl{ordinal width} strictly below
$\omegaOrd^2$, and a \kl{maximal order type} strictly below $\omegaOrd^3$
(\cref{ordinal-invariants-bounded:cor}).

In the case of \kl{downwards closed} languages, we prove that they are deeply
related to the notion of \kl{uniformly recurrent words}, borrowed from the
study of word combinatorics. As an intermediate result, we prove that an
infinite word $w$ is \kl{ultimately uniformly recurrent} if and only if the set
of all \kl{infixes} of $w$ is \kl{well-quasi-ordered} by the \kl{infix
relation} (\cref{ultimately-uniformly-recurrent:lem}). We show that every
language $L \subseteq \Sigma^*$ that is \kl{downwards closed} and
\kl{well-quasi-ordered} by the \kl{infix relation} is a finite union of the
sets of finite \kl{infixes} of some \kl{ultimately uniformly recurrent}
bi-infinite words. This also proves that such languages have an \kl{ordinal
height} of at most $\omegaOrd$, an \kl{ordinal width} strictly below
$\omegaOrd^2$, and a \kl{maximal order type} strictly below $\omegaOrd^3$
(\cref{small-ordinal-invariants:thm}).

Then, we turned our attention to decision procedures. To that end, we need to
choose a computational model representing languages. For \kl{downwards closed}
languages, because of their close connection with infinite words, we considered
a model based on \kl{automatic sequences}. Using this model, we can decide
whether a language is \kl{well-quasi-ordered} by the \kl{infix relation}
(\cref{automatic-wqo:thm}). We also studied another representation, where
languages are recognized by \kl{amalgamation systems} \cite{ASZZ24}. Such
systems will be formally introduced in \cref{amalgamation-systems:subsec}, but
for the moment let us just say that they include many classical computational
models such as finite automata, context-free grammars, and Petri nets
\cite{ASZZ24}. This provides us with a \emph{meta-algorithm} for deciding
whether a given language is \kl{well-quasi-ordered} by the \kl{prefix
relation}, the \kl{suffix relation}, or the \kl{infix relation} under mild
effective restrictions on the computational model that we call an \kl{effective
amalgamative class}, the formal definition of which we defer to
\cref{infixes-amalgamation-effective:subsec}. Given a class $\mathcal{C}$ that
is a \kl{strongly effective amalgamative class} of languages, we designed a
decision procedure that takes as input a language $L \in \mathcal{C}$, and
decides whether $L$ is \kl{well-quasi-ordered} by the \kl{prefix relation}, the
\kl{suffix relation}, or the \kl{infix relation}
(\cref{infix-wqo-is-emptiness:thm}). Quite surprisingly, we also showed that,
if a language recognized by an \kl{amalgamation system} is
\kl{well-quasi-ordered} for the \kl{infix relation}, then it is a \kl{bounded
language} (\cref{infix-amalgamation:thm}), which automatically bounds the
\kl{ordinal invariants} of the language. Let us point out that the above result
implies that the hypothesis of \kl{bounded languages} on the theoretical side
is not a restriction in practice. As a down-to-earth and more easily
understandable consequence, our generic decision procedure applies to the class
$\mathcal{C}_\text{aut}$ of all languages recognized by finite automata, and to
the class $\mathcal{C}_\text{cfg}$ of all languages recognized by context-free
grammars, which are both \kl{effective amalgamative classes}
(\cref{aut-cfg-infix:cor}).

Finally, we noticed that for \kl{downwards closed} languages that are
\kl{well-quasi-ordered} by the \kl{infix relation}, being
\kl(language){bounded} is the same as being \kl(language){regular}.
Furthermore, a \kl{bounded language} is \kl{well-quasi-ordered} by the
\kl{infix relation} if and only if its \kl{downwards closure} is
\kl{well-quasi-ordered} by the \kl{infix relation}
(\cref{bounded-wqo-dwclosed:cor}). This shows that, for \kl{bounded languages}
(and therefore, for all languages recognized by \kl{amalgamation systems}) that
are \kl{well-quasi-ordered} by the \kl{infix relation}, their \kl{downwards
closure} is a \kl{regular language}. This is a weak version of the usual result
that the \kl{downwards closure} for the \kl{scattered subword relation} is
always a \kl{regular language}.

\subparagraph{Related work} The study of alternative \kl{well-quasi-ordered}
relations over finite words is far from new. For instance, orders obtained by
so-called \emph{derivation relations} where already analysed by Bucher,
Ehrenfeucht, and Haussler \cite{BUEUD85}, and were later extended by
D'Alessandro and Varricchio \cite{ALVA03,ALVA06}. However, in all those cases
the orderings are \emph{multiplicative}, that is, if $u_1 \preceq v_1$ and $u_2
\preceq v_2$ then $u_1u_2 \preceq v_1v_2$. This assumption does not hold for
the \kl{prefix}, \kl{suffix}, and \kl{infix} relations.

\subparagraph{Outline} 
We introduce in \cref{prelims:sec} the
necessary background on \kl{well-quasi-orders} and \kl{ordinal invariants}.
In
\cref{prefixes:sec}, which is relatively
self-contained, we study the \kl{prefix relation} and prove in
\cref{prefixes:thm} the characterization of \kl{well-quasi-ordered}
languages by the \kl{prefix relation}. In
\cref{infixes-bounded:sec}, we
obtain the \kl[infix relation]{infix} analogue of \cref{prefixes:thm}
specifically for \kl{bounded languages}
(\cref{bounded-language:thm}). 
In \cref{infixes-dwclosed:sec}, we study the \kl{downwards closed}
languages, and compute bounds on their \kl{ordinal invariants} in \cref{small-ordinal-invariants:thm}.
Finally, 
we generalize these results to all
\kl{amalgamation systems} in \cref{infixes-amalgamation:sec}
in
(\cref{infix-amalgamation:thm}),
and provide a decision procedure for checking whether a language is
\kl{well-quasi-ordered} by the \kl{infix relation} (resp. \kl{prefix} and \kl{suffix}) in
this context (\cref{infix-wqo-is-emptiness:thm}).

\subparagraph{Acknowledgements} We would like to thank participants of the 2024
edition of \kl{Autobóz} for their helpful comments and discussions.
We would also like to thank Vincent Jugé for his pointers on word combinatorics.
\section{Preliminaries}
\label{prelims:sec}

\paragraph*{Finite words.}
\AP 
In this paper, we use letters $\Sigma, \Gamma$ to
denote finite alphabets, $\Sigma^*$ to denote the set of finite words over
$\Sigma$, and $\varepsilon$ for the empty word in $\Sigma^*$. In order to give
some intuition on the decision problems, we will sometimes use the notion of
\intro{finite automata}, \intro{regular languages}, and Monadic Second Order
logic ($\intro*\MSO$) over finite words, and assume is familiar with them. We
refer to the textbook of \cite{THOM97} for a detailed introduction to this
topic.

\paragraph*{Orderings and Well-Quasi-Orderings.}
\AP
A \intro{quasi-order} is a
reflexive and transitive binary relation, it is a \intro{partial order} if it
is furthermore antisymmetric. A \intro{total order} is a \kl{partial order}
where any two elements are comparable. Let now us introduce some notations for
\kl{well-quasi-orders}. A sequence $\seqof{x_i}$ in a set $X$ is
\intro(sequence){good} if there exist $i < j$ such that $x_i \leq x_j$. It is
\intro(sequence){bad} otherwise. Therefore, a \kl{well-quasi-ordered} set is a
set where every infinite sequence is \kl(sequence){good}. A \intro{decreasing
sequence} is a sequence $\seqof{x_i}$ such that $x_{i+1} < x_i$ for all $i$,
and an \intro{antichain} is a set of pairwise incomparable elements. An
equivalent definition of a \kl{well-quasi-ordered} set is that it contains no
infinite \kl{decreasing sequences}, nor infinite \kl{antichains}. We refer to
\cite{SCSC12} for a detailed survey on well-quasi-orders.

The \kl{prefix relation} (resp. the \kl{suffix relation} and the \kl{infix
relation}) on $\Sigma^*$ are always \intro{well-founded}, i.e., there are no
infinite \kl{decreasing sequences} for this ordering. In particular, for a
language $L \subseteq \Sigma^*$ to be \kl{well-quasi-ordered}, it suffices to
prove that it contains no infinite \kl{antichain}. 

\paragraph*{Ordinal Invariants.} \label{ordinal-invariants:subsec}
\AP
An \intro{ordinal} is a \kl{well-founded} \kl{totally ordered}
set. We use $\alpha, \beta, \gamma$ to denote ordinals, and use $\intro*\omegaOrd$ to
denote the first infinite \kl{ordinal}, i.e., the set of natural numbers with the
usual ordering. We also use $\intro*\omegaOne$ to denote the first \emph{uncountable}
ordinal.
We only assume superficial familiarity with ordinal arithmetic, and
refer to the books of Kunen \cite{KUNEN80} and Krivine~\cite[Chapter
II]{KRIVINE71} for a detailed introduction to this domain.
Given a tree $T$
whose branches are all finite we can define an \kl{ordinal} $\alpha_T$ inductively
as follows: if $T$ is a leaf then $\alpha_T = 0$, if $T$ has children
$\seqof{T_i}$ then $\alpha_T = \sup \setof{\alpha_{T_i} + 1}{i \in \Nat}$. We
say that $\alpha_T$ is the \emph{rank} of $T$. 

\AP
Let $(X, \leq)$ be a \kl{well-quasi-ordered} set. One can define three
well-founded trees from $X$: the tree of \kl{bad sequences}, the tree of
decreasing sequences, and the tree of \kl{antichains}. The rank of these
respective trees are called respectively the \intro{maximal order type} of $X$
written $\intro*{\oType{X}}$ \cite{dejongh77}, the \intro{ordinal height} of
$X$ written $\intro*{\oHeight{X}}$ \cite{schmidt81}, and the \intro{ordinal
width} of $X$ written $\intro*{\oWidth{X}}$ \cite{kriz90b}. These three
parameters are called the \intro{ordinal invariants} of a
\kl{well-quasi-ordered} set $X$. We refer to the survey of \cite{DZSCSC20} for
a detail discussion on these concepts and their computation on specific classes
of well-quasi-ordered sets.

\AP
We will use the following inequality between \kl{ordinal invariants}, due to
\cite{kriz90b}, and that was recalled in \cite[Theorem 3.8]{DZSCSC20}:
$\oType{X} \leq \oHeight{X} \oComProd \oWidth{X}$, where $\intro*\oComProd$ is
the \intro{commutative ordinal product}, also known as the \reintro{Hessenberg
product}. We will not recall the definition of this product here, and refer to
\cite[Section 3.5]{DZSCSC20} for a detailed introduction to this concept. The
only equalities we will use are $\omegaOrd \oComProd \omegaOrd = \omegaOrd^2$
and $\omegaOrd^2 \oComProd \omegaOrd = \omegaOrd^3$.

\section{Prefixes and Suffixes}
\label{prefixes:sec}

In this section, we study the well-quasi-ordering of languages under the
\kl{prefix relation}. Let us immediately remark that the map $u \mapsto u^R$
that reverses a word is an order-bijection between $(X^*, \prefleq)$ and $(X^*,
\suffleq)$, that is, $u \prefleq v$ if and only if $u^R \suffleq v^R$.
Therefore, we will focus on the prefix relation in the rest of this section, as
$(L, \prefleq)$ is \kl{well-quasi-ordered} if and only if $(L^R, \suffleq)$ is.

The next remark we make is that $\Sigma^*$ is not \kl{well-quasi-ordered} by
the \kl{prefix relation} as soon as $\Sigma$ contains two distinct letters $a$
and $b$. As an example of infinite \kl{antichain}, we can consider the set of
words $a^n b$ for $n \in \Nat$. As mentioned in the introduction, there are
however some languages  that are \kl{well-quasi-ordered} by the \kl{prefix
relation}. A simple example being the (regular) language $a^* \subseteq
\set{a,b}^*$, which is order-isomorphic to natural numbers with their usual
orderings $(\Nat, \leq)$.

In order to characterize the existence of infinite \kl{antichains} for the
\kl{prefix relation}, we will introduce the following tree.

\begin{figure}
    \centering
    \begin{tikzpicture}[
        isWord/.style={rectangle,inner sep=0mm},
        isEmbedding/.style={->, >=stealth, thick},
        branch/.style={A5},
        antichain/.style={A2},
        branchEdge/.style={A5,dashed},
        outEdge/.style={dotted,B2},
        isInfixEnc/.style={},
        isRealWord/.style={A4,opacity=0.2},
        isRealEmbedding/.style={A4,opacity=0.2},
        isNewEdge/.style={B2,dashed},
        ]

\node[branch,isWord] (epsilon) at (-0.5, 0) {\strut $\varepsilon$};
\node[branch,isWord] (a) at (-1.0, 1) {\strut a};
\node[antichain,isWord] (b) at (0.0, 1) {\strut b};
\node[branch,isWord] (aa) at (-2.0, 2) {\strut aa};
\node[antichain,isWord] (ab) at (-1.0, 2) {\strut ab};
\node[isWord] (ba) at (0.0, 2) {\strut ba};
\node[isWord] (bb) at (1.0, 2) {\strut bb};
\node[branch,isWord] (aaa) at (-4.0, 3) {\strut aaa};
\node[antichain,isWord] (aab) at (-3.0, 3) {\strut aab};
\node[isWord] (aba) at (-2.0, 3) {\strut aba};
\node[isWord] (abb) at (-1.0, 3) {\strut abb};
\node[isWord] (baa) at (0.0, 3) {\strut baa};
\node[isWord] (bab) at (1.0, 3) {\strut bab};
\node[isWord] (bba) at (2.0, 3) {\strut bba};
\node[isWord] (bbb) at (3.0, 3) {\strut bbb};
\draw[branchEdge,isEmbedding] (epsilon) -- (a);
\draw[outEdge,isEmbedding] (epsilon) -- (b);
\draw[branchEdge,isEmbedding] (a) -- (aa);
\draw[outEdge,isEmbedding] (a) -- (ab);
\draw[isEmbedding] (b) -- (ba);
\draw[isEmbedding] (b) -- (bb);
\draw[branchEdge,isEmbedding] (aa) -- (aaa);
\draw[outEdge,isEmbedding] (aa) -- (aab);
\draw[isEmbedding] (ab) -- (aba);
\draw[isEmbedding] (ab) -- (abb);
\draw[isEmbedding] (ba) -- (baa);
\draw[isEmbedding] (ba) -- (bab);
\draw[isEmbedding] (bb) -- (bba);
\draw[isEmbedding] (bb) -- (bbb);

\end{tikzpicture}
    \caption{An \kl{antichain branch} for the language $a^* b$,
        represented in the \kl{tree of prefixes} over the alphabet $\set{a,b}$.
        The branch is represented with dashed lines in turquoise, and the
        \kl{antichain} is represented in dotted lines in blood-red.
    }
    \label{antichain-branch:fig}
\end{figure}
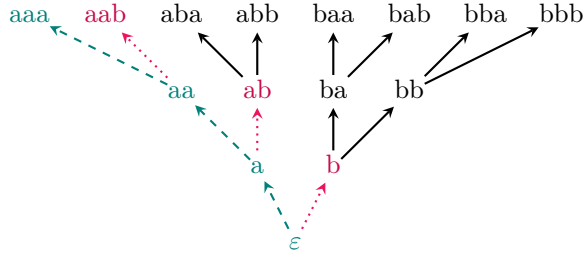

\begin{definition}
    The \intro{tree of prefixes} over a finite alphabet $\Sigma$
    is the infinite tree $T$ whose nodes are the words of $\Sigma^*$, and
    such that the children of a word $w$ are the words $wa$ for all $a \in
    \Sigma$. 
\end{definition}

We will use this \kl{tree of prefixes} to find simple witnesses
of the existence of infinite \kl{antichains} in the \kl{prefix relation}
for a given language $L$, namely by introducing \kl{antichain branches}.

\begin{definition}
    An \intro{antichain branch} for a language $L$ is an infinite 
    branch $B$ of the \kl{tree of prefixes} such that from every point of the branch, 
    one can reach a word in $L \setminus B$. Formally:
    $\forall u \in B, \exists v \in \Sigma^*, uv \in L \setminus B$.
\end{definition}

Let us illustrate the notion of \kl{antichain branch} over the alphabet $\Sigma
= \set{a,b}$, and the language $L = a^* b$. In this case, the set $a^*$ (which
is a branch of the \kl{tree of prefixes}) is an \kl{antichain branch} for $L$.
This holds because for any $a^k$, the word $a^k \prefleq a^kb$ belongs to $L
\setminus a^*$. In general, the existence of an \kl{antichain branch} for a
language $L$ implies that $L$ contains an infinite \kl{antichain}, and because
the alphabet $\Sigma$ is assumed to be finite, one can leverage the fact that
the \kl{tree of prefixes} is finitely branching to prove that the converse
holds as well.

\begin{lemma}
    \label{antichain-branches-prefix:lem}
    Let $L \subseteq \Sigma^*$ be a language. Then, $L$ contains an infinite
    \kl{antichain} if and only if there exists an \kl{antichain branch} for $L$.
\end{lemma}
\begin{proof}
    Assume that $L$ contains an \kl{antichain branch}. Let us construct an
    infinite \kl{antichain} as follows. We start with a set $A_0 \defined
    \emptyset$ and a node $v_0$ at the root of the tree. At step $i$, we
    consider a word $w_i$ such that $v_i$ is a \kl{prefix} of $w_i$, and $w_i
    \in L \setminus B$, which exists by definition of \kl{antichain branches}.
    We then set $A_{i+1} \defined A_i \cup \set{w_i}$. To compute $v_{i+1}$, we
    consider the largest prefix of $w_i$ that belongs to $B$, and set $v_{i+1}$
    to be the successor of this prefix in $B$. By an immediate induction, we
    conclude that for all $i \in \Nat$, $A_i$ is an \kl{antichain}, and that
    $v_i$ is a node in the \kl{antichain branch} $B$ such that $v_i$ is not a
    prefix of any word in $A_i$. 

    Conversely, assume that $L$ contains an infinite \kl{antichain} $A$. Let us
    construct an \kl{antichain branch}. Let us consider the subtree of the
    \kl{tree of prefixes} that consists in words that are \kl{prefixes} of
    words in $A$. This subtree is infinite, and by König's lemma, it contains
    an infinite branch. By definition this is an \kl{antichain branch}.
\end{proof}

One immediate application of \cref{antichain-branches-prefix:lem} is
that \kl{antichain branches} can be described inside the \kl{tree of prefixes}
by a monadic second order formula ($\MSO$-formula), allowing us to
leverage the decidability of $\MSO$ over infinite binary trees \cite[Theorem
1.1]{RAB69}. This result will follow from our general decidability result
(\cref{infix-wqo-is-emptiness:thm}) but is worth stating on its own for its
simplicity.

\begin{corollary}
    \label{prefix-wqo-reg-decidable:cor}
    If $L$ is regular, then the existence of an infinite antichain is decidable.
\end{corollary}
\begin{proof}
    If $L$ is regular, then it is $\MSO$-definable, and there 
    exists a formula $\varphi(x)$ in $\MSO$ that selects nodes 
    of the \kl{tree of prefixes} that belong to $L$. Now, to decide whether there
    exists an \kl{antichain branch} for $L$, we can simply check whether
    the following formula is satisfied:
    \begin{equation*}
        \exists B. 
        B \text{ is a branch } \land
        \forall x \in B, \exists y. y \text{ is a child of } x \land \varphi(y) \land y \not\in B
        \quad .
    \end{equation*}
    Because the above formula is an $\MSO$-formula over the infinite
    $\Sigma$-branching tree, whether it is satisfied is decidable
    as an easy consequence of the decidability of $\MSO$ over infinite binary
    trees
    \cite[Theorem 1.1]{RAB69}.
\end{proof}

Let us now go further and fully characterize languages $L$ such that the
prefix relation is well-quasi-ordered, without any restriction on the
decidability of $L$ itself. Let us remark that finite unions of \kl{chains} are
always \kl{well-quasi-ordered} by the \kl{prefix relation} because they lack
infinite \kl{antichains} by definition. The following theorem states that this
is the only possible reason for a language $L$ to be \kl{well-quasi-ordered} by
the \kl{prefix relation}.

For the proof of the following theorem, we will use special notations to
describe the \intro{upwards closure} of a set $S$ for a relation $\preceq$,
which is defined as $\upset[\preceq]{S} \defined \setof{y \in \Sigma^*}{
\exists x \in S. x \preceq y}$. Anticipating the use of the symmetric notion of
\intro{downwards closure}, let us introduce it as follows: $\dwset[\preceq]{S} \defined \setof{y
\in \Sigma^*}{ \exists x \in S. y \preceq x}$. Abusing notations, we will
write $\upset{w}$ and $\dwset{w}$ for the upwards and downwards closure of a
single element $w$, omitting the ordering relation when it is clear from the
context. A set $S$ is called \intro{downwards closed} if $\dwset{S} = S$.

\begin{theorem}
    \label{prefixes:thm}
    A language $L \subseteq \Sigma^*$ is \kl{well-quasi-ordered} by the
    \kl{prefix relation} if and only if $L$ is a union of \kl{chains}.
\end{theorem}
\begin{proof}
    Assume that $L$ is a finite union of \kl{chains}.
    Because the \kl{prefix relation} is \kl{well-founded},
    and that finite unions of \kl{chains} have finite \kl{antichains},
    we conclude that $L$ is \kl{well-quasi-ordered}.

    Conversely, assume that $L$ is \kl{well-quasi-ordered} by the \kl{prefix
    relation}. Let us define $S$ the set of words $w$ such that there exists
    two words $wu$ and $wv$ both in $L$ that are not comparable for the
    \kl{prefix relation}. Assume by contradiction that $S$ is infinite.
    Then, $S$ equipped with the \kl{prefix relation} is an infinite
    tree with finite branching, and therefore contains an infinite
    branch, which is by definition an \kl{antichain branch} for $L$.
    This contradicts the assumption that $L$ is \kl{well-quasi-ordered}.
    Now, consider $w$ be a maximal element for the \kl{prefix ordering}
    in $S$. By definition, all words in $L$ that contain $w$ as a prefix
    must be comparable for the \kl{prefix relation}. Therefore,
    $(\upset[\prefleq]{w}) \cap L$ is a \kl{chain} for the \kl{prefix relation}.
    In particular, letting $S_{\max}$ be the set of all maximal elements
    of $S$,
    we conclude that 
    \begin{equation*}
        L \subseteq S \cup \bigcup_{w \in S_{\max}} (\upset[\prefleq]{w}) \cap L
        \quad .
    \end{equation*}
    Hence, that $L$ is finite union of \kl{chains}.
\end{proof}

As an immediate consequence, we have a very fine-grained understanding of the
\kl{ordinal invariants} of such \kl{well-quasi-ordered} languages, which can be
leveraged in bounding the complexity of algorithms working on such languages.

\begin{corollary}
    \label{prefixes-ordinal-invariants:cor}
    Let $L \subseteq \Sigma^*$ be a language that is
    \kl{well-quasi-ordered} by the \kl{prefix relation}. Then,
    \kl{maximal order type} of $L$ strictly smaller than $\omega^2$,
    the \kl{ordinal height} of $L$ is at most $\omega$, and
    its \kl{ordinal width} is finite.
\end{corollary}
\section{Infixes and Bounded Languages}
\label{infixes-bounded:sec}

\AP In this section, we study languages equipped with the \kl{infix relation}.
As opposed to the \kl{prefix} and \kl{suffix} relations, the \kl{infix
relation} can lead to very complicated \kl{well-quasi-ordered} languages.
Formally, the upcoming \cref{infix-embedding:thm} due to Kuske shows that
\emph{any} countable partial-ordering with finite initial segments can be
embedded into the infix relation of a language. To make the former statement
precise, let us recall that an \intro{order embedding} from a quasi-ordered set
$(X, \preceq)$ into a quasi-ordered set $(Y, \preceq')$ is a function $f \colon
X \to Y$ such that for all $x, y \in X$, $x \preceq y$ if and only if $f(x)
\preceq' f(y)$. When such an embedding exists, we say that $X$ \reintro{embeds
into} $Y$. Recall that a quasi-ordered set $(X, \preceq)$ is a \kl{partial
ordering} whenever the relation $\preceq$ is antisymmetric, that is $x \preceq
y$ and $y \preceq x$ implies $x = y$. 
A simplified version of the embedding defined in \cref{infix-embedding:thm} is illustrated
for the \kl{subword relation} in \cref{infix-embedding:fig}.
\begin{lemma}{\cite[Lemma 5.1]{DBLP:journals/ita/Kuske06}}
    \label{infix-embedding:thm}
    Let $(X, \preceq)$ be a \kl{partially ordered} set,
    and $\Sigma$ be an alphabet with at least two letters.
    Then the following are equivalent:
    \begin{enumerate}
        \item 
            $X$ \kl{embeds into} $(\Sigma^*, \infleq)$,
        \item 
            $X$ is countable, and for every $x \in X$,
            its \kl{downwards closure}
            $\dwset[\preceq]{x}$ is finite.
    \end{enumerate}
\end{lemma}
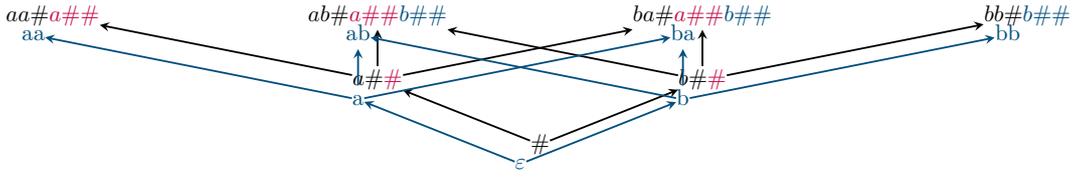
\begin{figure}
    \centering
    \begin{tikzpicture}[
        isWord/.style={rectangle,inner sep=0mm},
        isEmbedding/.style={->, >=stealth, thick},
        branch/.style={A5},
        antichain/.style={A2},
        branchEdge/.style={A5,dashed},
        outEdge/.style={dotted,B2},
        isInfixEnc/.style={},
        isRealWord/.style={A4,opacity=0.2},
        isRealEmbedding/.style={A4,opacity=0.2},
        isNewEdge/.style={B2},
        ]

\newcommand{\sep}[1]{\textcolor{gray}{\#^{#1}}}
\node[isWord,isInfixEnc] (Iepsilon) at (-4.7, 0.3) {\strut $\#$};
\node[isWord,isInfixEnc] (Ia) at (-7.2, 1.3) {\strut $a\#\textcolor{A2}{\#}$};
\node[isWord,isInfixEnc] (Ib) at (-2.2, 1.3) {\strut $b\#\textcolor{A2}{\#}$};
\node[isWord,isInfixEnc] (Iaa) at (-12.2, 2.3) {\strut $aa\#\textcolor{A2}{a\#\#}$};
\node[isWord,isInfixEnc] (Iab) at (-7.2, 2.3) {\strut $ab\#\textcolor{A2}{a\#\#}\textcolor{A4}{b\#\#}$};
\node[isWord,isInfixEnc] (Iba) at (-2.2, 2.3) {\strut $ba\#\textcolor{A2}{a\#\#}\textcolor{A4}{b\#\#}$};
\node[isWord,isInfixEnc] (Ibb) at (2.8, 2.3) {\strut $bb\#\textcolor{A4}{b\#\#}$};
\draw[isEmbedding] (Iepsilon) -- (Ia);
\draw[isEmbedding] (Iepsilon) -- (Ib);
\draw[isEmbedding] (Ia) -- (Iaa);
\draw[isEmbedding] (Ia) -- (Iab);
\draw[isEmbedding] (Ia) -- (Iba);
\draw[isEmbedding] (Ib) -- (Iab);
\draw[isEmbedding] (Ib) -- (Iba);
\draw[isEmbedding] (Ib) -- (Ibb);
\node[isWord,isInfixEnc,isRealWord] (epsilon) at (-5.0, 0) {\strut $\varepsilon$};
\node[isWord,isInfixEnc,isRealWord] (a) at (-7.5, 1) {\strut a};
\node[isWord,isInfixEnc,isRealWord] (b) at (-2.5, 1) {\strut b};
\node[isWord,isInfixEnc,isRealWord] (aa) at (-12.5, 2) {\strut aa};
\node[isWord,isInfixEnc,isRealWord] (ab) at (-7.5, 2) {\strut ab};
\node[isWord,isInfixEnc,isRealWord] (ba) at (-2.5, 2) {\strut ba};
\node[isWord,isInfixEnc,isRealWord] (bb) at (2.5, 2) {\strut bb};
\draw[isEmbedding,isRealEmbedding] (epsilon) -- (a);
\draw[isEmbedding,isRealEmbedding] (epsilon) -- (b);
\draw[isEmbedding,isRealEmbedding] (a) -- (aa);
\draw[isEmbedding,isRealEmbedding] (a) -- (ab);
\draw[isEmbedding,isRealEmbedding] (a) -- (ba);
\draw[isEmbedding,isRealEmbedding] (b) -- (ab);
\draw[isEmbedding,isRealEmbedding] (b) -- (ba);
\draw[isEmbedding,isRealEmbedding] (b) -- (bb);

\end{tikzpicture}
    \caption{Representation of the \kl{subword relation} for $\set{a,b}^*$
        inside the \kl{infix relation} for $\set{a,b,\#}^*$
        using a simplified version of \cref{infix-embedding:thm}, restricted to words
        of length at most $3$. 
    }
    \label{infix-embedding:fig}
\end{figure}

\AP As a consequence of \cref{infix-embedding:thm}, we cannot replay
proofs of \cref{prefixes:sec}, and will
actually need to leverage some regularity of the languages to obtain a
characterization of \kl{well-quasi-ordered} languages under the \kl{infix
relation}. This regularity will be imposed through the notion of \intro{bounded
languages}, i.e., languages $L \subseteq \Sigma^*$ such that there exists words
$w_1, \dots, w_n$ satisfying $L \subseteq w_1^* \cdots w_n^*$.

\begin{theorem}[restate=bounded-language:thm,label=bounded-language:thm]
    Let $L$ be a \kl{bounded language} of $\Sigma^*$. Then,
    $L$ is a \kl{well-quasi-order} when endowed with the 
    \kl{infix relation} if and only if it included in a finite union of 
    products $S_i \cdot P_i$ where 
    $S_i$ is a \kl{chain} for the \kl{suffix relation}, and 
    $P_i$ is a \kl{chain} for the \kl{prefix relation},
    for all $1 \leq i \leq n$.
\end{theorem}

Let us first remark that if $S$ is a \kl{chain} for the \kl{suffix relation}
and $P$ is a \kl{chain} for the \kl{prefix relation}, then $SP$ is
\kl{well-quasi-ordered} for the \kl{infix relation}. This proves the (easy)
right-to-left implication of \cref{bounded-language:thm}. Furthermore, any
language $L$ included in a finite union of products $S_i \cdot P_i$ where $S_i$
is in fact a \kl{bounded language}.

\AP In order to prove the (difficult) left-to-right implication of
\cref{bounded-language:thm}, we will rely heavily on the
combinatorics of periodic words. Let us recall that a non-empty word $w \in
\Sigma^+$ is \intro(word){periodic} with period $x \in \Sigma^*$ if there
exists a $p \in \Nat$ such that $w \infleq x^p$. The \intro{periodic length} of
a word $u$ is the minimal length of a word $x$ such that $u$ is an \kl{infix}
of $x^p$ for some $p \in \Nat$ and $x \in \Sigma^+$.

\begin{lemma}
    \label{periodic-infixes:lem}
    Let $u,v \in \Sigma^*$ be two (non-empty) \kl{periodic words}
    having \kl{periodic lengths} $p$ and $q$ respectively.
    Then, if $u \infleq v$ and $\card{u} \geq \factorial[p]{p \times q}$,
    then $u$ and $v$ share the same \kl{periodic length}
    $p = q$.
\end{lemma}
\begin{proof}
    The fact that $u$ and $v$ are \kl{periodic length}
    respectively $p$ and $q$ translates into the fact that $u_{i+p} = u_i$ and
    $v_{i+q} = v_i$ for all indices $i \in \Nat$ such that those letters are
    well-defined.

    Now, assume that $u$ is an \kl{infix} of $v$, this provides the existence
    of a $k \in \Nat$ such that $u = v_{k} \cdots v_{k + \card{u} - 1}$. In
    particular, $v_{k+i+p} = v_{k+i}$ for all $i \in \Nat$ such that $k+i+p < k
    + \card{u}$. Since we also have $v_{k+i+q} = v_{k+i}$ for all $1 \leq i
    \leq \card{v} - k - q$. We conclude that both $u$ and $v$ are of
    \kl{periodic length} the greatest common divisor of $p$ and $q$, and by
    minimality of $q$ this must be equal to $q$ and to $p$.
\end{proof}

\begin{corollary}
    \label{powers-infixes:cor}
    Let $u,v \in \Sigma^*$ and $k, \ell \in \Nat$
    such that $k \geq \factorial[p]{\card{u} \times \card{v}}$,
    $\ell \geq \factorial[p]{\card{v} \times \card{u}}$,
    and $u^k \infleq v^\ell$.
    Then, there exists $w \in \Sigma^*$ of size at most
    $\min \set{\card{u}, \card{v}}$ and a $p \in \Nat$
    such that
    $u^k \infleq v^\ell \infleq w^p$.
\end{corollary}

The reason why \kl{periodic words} built using a given period $x \in \Sigma^+$
are interesting for the \kl{infix relation} is that they naturally create
\kl{chains} for the \kl{prefix} and \kl{suffix} relations. Indeed, if $x \in
\Sigma^+$ is a finite word, then $\setof{x^p}{p \in \Nat}$ is a \kl{chain} for
the \kl{infix relation}. Note that in general, the \kl{downwards closure} of a
chain is \emph{not} a chain (see \cref{dw-closure-not-wqo:rem}). However, for the chains generated using periodic
words, the \kl{downwards closure} $\dwset[\infleq]{\setof{x^p}{p \in \Nat}}$ is
a \emph{finite union} of \kl{chains}. Because this set will appear in bigger
equations, we introduce the shorter notation $\intro*\InfPeriodChain{x}$ for
the set of \kl{infixes} of words of the form $x^p$, where $p$ ranges over
$\Nat$.

\begin{remark}
    \label{dw-closure-not-wqo:rem}
    Let $(X,\preceq)$ be a quasi-ordered set, and $L \subseteq X$ be such that $(L,
    \preceq)$ is \kl{well-quasi-ordered}. It is not true in general that
    $(\dwset{L}, \preceq)$ is \kl{well-quasi-ordered}. In the case of $(\Sigma^*,
    \infleq)$ a typical example is to start from an infinite \kl{antichain} $A$,
    together with an enumeration $\seqof{w_i}[i \in \Nat]$ of $A$, and build the language $L
    \defined \setof{ \prod_{i = 0}^n w_i }{ i \in \Nat }$. By definition, $L$ is a
    \kl{chain} for the \kl{infix} ordering, hence \kl{well-quasi-ordered}. However,
    $\dwset[\infleq]{L}$ contains $A$, and is therefore not
    \kl{well-quasi-ordered}. 
\end{remark}

\begin{lemma}
    \label{inf-period-chain:lem}
    Let $x \in \Sigma^+$ be a word, and
    Then $\InfPeriodChain{x}$ is a finite union of \kl{chains}
    for the \kl{infix}, \kl{prefix} and \kl{suffix} relations 
    simultaneously.
\end{lemma}
\begin{proof}
    Let $x \in \Sigma^+$ be a word, and let $P_x$ be the (finite) set 
    of all \kl{prefixes} of $x$, and $S_x$ be the (finite)
    set of all \kl{suffixes} of $x$.
    Assume that $w \in \InfPeriodChain{x}$, then $w = u x^p v$ for some
    $u \in S_x$, $v \in P_x$, and $p \in \Nat$.
    We have proven that
    \begin{equation*}
        \InfPeriodChain{x} \subseteq \bigcup_{u \in P_x} \bigcup_{v \in S_x} u x^* v
        \quad .
    \end{equation*}

    Let us now demonstrate that for all $(u,v) \in S_x \times P_x$, the
    language $u x^* v$ is a \kl{chain} for the \kl{infix}, \kl{suffix} and \kl{prefix} relations.
    To that end,
    let $(u,v) \in S_x \times P_x$ and $\ell, k \in \Nat$ be such that $\ell <
    k$, let us prove that $u x^\ell v \infleq u x^k  v$. Because $v \prefleq
    x$, we know that there exists $w$ such that $vw = x$. In particular,
    $ux^\ell vw = u x^{\ell + 1}$, and because $\ell < k$, we conclude that $u
    x^{\ell + 1} \prefleq u x^k v$. By transitivity, $u x^\ell v \prefleq u x^k
    v$, and \emph{a fortiori}, $u x^\ell v \infleq u x^k v$. 
    Similarly, because $u \suffleq x$,  there exists $w$ such that $wu  = x$, 
    and we conclude that $u x^{\ell} v \suffleq w u x^\ell v = x^{\ell + 1} v \suffleq u x^k v$.
    \qedhere
\end{proof}

The following combinatorial lemma connects the property of being
\kl{well-quasi-ordered} to a property of the \kl{periodic lengths} of words in
a language, based on the assumption that some factors can be iterated. It is
the core result that powers the analysis done in the upcoming
\cref{bounded-language:thm,infix-amalgamation:thm}.

\begin{lemma}
    \label{pumping-periods:lem}
    Let $L \subseteq \Sigma^*$ be a language
    that is \kl{well-quasi-ordered} by the \kl{infix relation}.
    Let $k \in \Nat$, $u_1, \cdots, u_{k+1} \in \Sigma^*$,
    and $v_1, \cdots, v_{k} \in \Sigma^+$
    be such that
    $w[\vec{n}] \defined (\prod_{i = 1}^k u_i v_i^{n_i}) u_{k+1}$
    belongs to $L$
    for arbitrarily large values of $\vec{n} \in \Nat^k$.
    Then, 
    there exists $x,y \in \Sigma^+$ of size 
    at most $\max \setof{\card{v_i}}{1 \leq i \leq k}$
    such that 
    one of the following holds for all
    $\vec{n} \in \Nat^{k}$:
    \begin{enumerate}
        \item $w[\vec{n}] \in u_1 \InfPeriodChain{x}$,
        \item $w[\vec{n}] \in \InfPeriodChain{x} u_{k+1}$,
        \item $w[\vec{n}] \in \InfPeriodChain{x} u_i \InfPeriodChain{y}$
            for some $1 \leq i \leq k + 1$.

    \end{enumerate}
\end{lemma}
\begin{proof}
    Note that the result is obvious if $k = 0$, and therefore
    we assume $k \geq 1$ in the following proof.

    Let us construct a sequence of words $\seqof{w_i}[i \in \Nat]$, where $w_i
    \defined w[\vec{n_i}]$ for some well-chosen indices $\vec{n_i} \in \Nat^k$. The goal
    being that 
    if $w[\vec{n_i}]$ is an \kl{infix} of $w[\vec{n_j}]$,
    then it can intersect at most \emph{two} iterated words,
    with an intersection that is long enough to successfully apply
    \cref{periodic-infixes:lem}.
    In order to achieve this,
    let us first define $s$ as the maximal size of a word $v_i$
    ($1 \leq i \leq k$) and $u_j$ ($1 \leq j \leq k+1$).
    Then,
    we consider $\vec{n_0} \in \Nat^k$ such that $\vec{n_0}$ has all 
    its components greater than $\factorial{s}$ and such that
    $w[\vec{n_0}]$ belongs to $L$. 
    Then, we inductively define 
    $\vec{n_{i+1}}$  as the smallest vector of numbers greater than $\vec{n_i}$,
    such that $w[\vec{n_{i+1}}]$ belongs to $L$, 
    and with $\vec{n_i}$ having all components greater than
    $2\card{w[\vec{n_i}]}$.

    Let us assume that $k \geq 2$ in the following proof for symmetry purposes,
    and argue later on that when $k = 1$ the same argument goes through.
    Because $L$ is \kl{well-quasi-ordered} by the \kl{infix relation}, there
    exists $i < j$ such that $w[\vec{n_i}]$ is an \kl{infix} of $w[\vec{n_j}]$.
    Now, because of the chosen values for $\vec{n_j}$, there exists $1 \leq \ell \leq
    k-1$ such that $w[\vec{n_i}]$ is actually an \kl{infix} of $u_{\ell}
    v_{\ell}^{n_{j,\ell}} u_{\ell+1} v_{\ell+1}^{n_{j,\ell+1}} u_{\ell+2}$.
    Even more,
    one of the three following equations holds:
    \begin{itemize}
        \item $w[\vec{n_i}] \infleq v_{\ell}^{n_{j,\ell}} u_{\ell+1} v_{\ell+1}^{n_{j,\ell+1}}$,
        \item $w[\vec{n_i}] \infleq u_{\ell}
            v_{\ell}^{n_{j,\ell}}$,
        \item $w[\vec{n_i}] \infleq
            v_{\ell+1}^{n_{j,\ell+1}} u_{\ell+2}$.
    \end{itemize}
    In all those cases, we conclude using \cref{powers-infixes:cor}
    that there exists $x,y \in \Sigma^+$ of size at most $s$, and 
    a number $1 \leq t \leq k$ such that
    $v_i^{n_i} \in \InfPeriodChain{x}$ for all $1 \leq i \leq t$,
    and
    $v_i^{n_i} \in \InfPeriodChain{y}$ for all $t < i \leq k$.
    In particular,
    $w[\vec{n_i}] \in \InfPeriodChain{x} u_{t} \InfPeriodChain{y}$.

    When $k = 1$, the situation is a bit more specific since we only have two
    cases: either $w_i \infleq u_1 v_1^{n_j}$ or $w_i \infleq v_1^{n_j} u_2$,
    and we conclude with an identical reasoning.
\end{proof}

\begin{lemma}
    \label{bounded-language:lem}
    Let $L \subseteq \Sigma^*$ be a \kl{bounded language}
    that is \kl{well-quasi-ordered} by the \kl{infix relation}.
    Then, there exists a finite subset $E \subseteq (\Sigma^*)^3$,
    such that:
    \begin{equation*}
        L \subseteq \bigcup_{(x,u,y) \in E} \InfPeriodChain{x} u \InfPeriodChain{y}
        \quad .
    \end{equation*}
\end{lemma}
\begin{proof}
    Let $w_1, \dots, w_n$ be such that
    $L \subseteq w_1^* \cdots w_n^*$.
    Let us define $m \defined \max \setof{\card{w_i}}{1 \leq i \leq n}$

    Let $w[\vec{k}] \defined w_1^{k_1} \cdots w_n^{k_n}$ be a map from $\Nat^k$
    to $\Sigma^*$. We are interested in the intersection of the image of $w$
    with $L$. Let us assume for instance that for all $\vec{k} \in \Nat^n$,
    there exists $\vec{\ell} \geq \vec{k}$ such that $w[\vec{\ell}] \in L$.
    Then, leveraging \cref{pumping-periods:lem}, we conclude that there exists
    $x,y$ of size at most $\max\setof{\card{w_i}}{1 \leq i \leq n}$ such that
    $w[\vec{k}] \in \InfPeriodChain{x} \cup \InfPeriodChain{x}
    \InfPeriodChain{y}$, and we conclude that $L \subseteq \InfPeriodChain{x}
    \cup \InfPeriodChain{x} \InfPeriodChain{y}$.

    Now, it may be the case that one cannot simultaneously assume that two
    component of the vector $\vec{k}$ are unbounded. In general, given a set $S
    \subseteq \set{1, \dots, n}$ of indices, we say that $S$ is admissible if
    there exists a bound $N_0$ such that for all $\vec{b} \in \Nat^S$, there
    exists a vector $\vec{k} \in \Nat^n$, such that $\vec{k}$ is greater than
    $\vec{b}$ on the $S$ components, and the other components are below the
    bound $N_0$. The language of an admissible set $S$ is the set of words
    obtained by repeating $w_i$ at most $N_0$ times if it is not in $S$
    ($w_i^{\leq N_0}$) and arbitrarily many times otherwise ($w_i^*$).
    Note that $L \subseteq \bigcup_{S \text{ admissible }} L(S)$.

    Now, admissible languages are ready to be pumped according to
    \cref{pumping-periods:lem}. For every admissible language,
    the size of a word that is not iterated is at most
    $N_0 \times m$ by definition, and we conclude that:
    \begin{equation}
        \label{bounded-language:eq}
        L \subseteq 
        \bigcup_{x,y \in \Sigma^{\leq n}}
        \bigcup_{u \in \Sigma^{\leq m \times N_0}}
        \InfPeriodChain{x} u \InfPeriodChain{y}
        \cup
        \InfPeriodChain{x} u
        \cup
        u \InfPeriodChain{x}
        \quad .
    \end{equation}
\end{proof}

\begin{proofof}{bounded-language:thm}
    We apply \cref{bounded-language:lem}, and conclude
    because $\InfPeriodChain{x}$ is a finite union of \kl{chains}
    for the \kl{prefix}, \kl{suffix} and \kl{infix} relations
    (\cref{inf-period-chain:lem}).
\end{proofof}

Let us now discuss the implications of this characterization in terms of
\kl{downwards closures}: if $L$ is a \kl{bounded language}, then considering
$L$ or its \kl{downwards closure} is equivalent with respect to being
\kl{well-quasi-ordered} by the \kl{infix relation}.

\begin{corollary}
    \label{bounded-wqo-dwclosed:cor}
    Let $L$ be a \kl{bounded language} of $\Sigma^*$. Then,
    $L$ is a \kl{well-quasi-order} when endowed with the
    \kl{infix relation} if and only if $\dwset[\infleq]{L}$ is.
\end{corollary}
\begin{proof}
    Because $L \subseteq \dwset[\infleq]{L}$, the right-to-left implication
    is trivial.
    For the left-to-right implication, let us assume that $L$ is a
    \kl{well-quasi-ordered} language for the \kl{infix relation}.
    Then $L$ is included in a finite union 
    of products of \kl{chains}:
    \begin{equation*}
        L \subseteq \bigcup_{i = 1}^n S_i \cdot P_i \quad .
    \end{equation*}
    Remark that the \kl{downwards closure} of a product of two \kl{chains}
    is a finite union of products of two chains.
    As a consequence, we conclude that $\dwset[\infleq]{L}$ is itself included
    in a finite union of products of \kl{chains}.
    Note that this also proves that $\dwset[\infleq]{L}$ is a \kl{bounded language},
    hence that it is \kl{well-quasi-ordered} by the \kl{infix relation} 
    via
    \cref{bounded-language:thm}.
\end{proof}

\begin{corollary}
    \label{ordinal-invariants-bounded:cor}
    Let $L$ be a \kl{bounded language} of $\Sigma^*$
    that is \kl{well-quasi-ordered} by the \kl{infix relation}.
    Then, the \kl{ordinal width} of $L$ is finite,
    its \kl{ordinal height} is at most $\omega$,
    and its \kl{maximal order type} is at most $\omega$.
\end{corollary}

\section{Infixes and Downwards Closed Languages}
\label{infixes-dwclosed:sec}

One may think that all \kl{downwards closed} languages for the \kl{infix
relation} that are \kl{well-quasi-ordered} are \kl(language){bounded}. Note
that this is what happens in the case of the \kl{subword embedding}, where any
\kl{downwards closed} language for this relation is characterized by finitely
many excluded \kl{subwords}, hence provides a \kl{bounded language}. However, this
is not the case for the \kl{infix relation}, as we will now illustrate with the
following two examples.

\begin{example}
    \label{dwclosed-wqo-not-finite-excl:ex}
    Let $L \defined a^* b^* \cup b^* a^*$. This language is \kl{downwards
    closed} for the \kl{infix relation}, is \kl{well-quasi-ordered} for the
    \kl{infix relation}, but is characterized by an \emph{infinite} number 
    of excluded infixes, respectively of the form $ab^ka$ and $ba^kb$ where $k \geq 1$.
\end{example}

To strengthen \cref{dwclosed-wqo-not-finite-excl:ex}, we will
leverage the \intro{Thue-Morse sequence} $\intro*\ThueMorse \in
\set{0,1}^{\Nat}$, which we will use as a black-box for its two main
characteristics: it is \kl{cube-free} and \kl{uniformly recurrent}. Being
\intro{cube-free} means that no (finite) word of the form $uuu$ is an
\kl{infix} of $\ThueMorse$, and being \intro{uniformly recurrent} means that
for every word $u$ that is an \kl{infix} of $\ThueMorse$, there exists $k \geq
1$ such that $u$ is an \kl{infix} of every word $v$ of size at least $k$. We
refer the reader to a nice survey of Allouche and Shallit for more information
on this sequence and its properties \cite{ALSHA99}.

\begin{lemma}
    \label{uniformly-recurrent:lem}
    Let $w \in \Sigma^\Nat$ be a \kl{uniformly recurrent} word.
    Then, the set of finite \kl{infixes} of $w$ is \kl{well-quasi-ordered} for the \kl{infix relation}.
\end{lemma}
\begin{proof}
    Let $L$ be the set of finite \kl{infixes} of $w$.
    Consider a sequence $\seqof{u_i}[i \in \Nat]$ of words in $L$. Without loss of
    generality, we may consider a subsequence such that $\card{u_i} <
    \card{u_{i+1}}$ for all $i \in \Nat$. Because $\ThueMorse$ is \kl{uniformly
    recurrent}, there exists $k \geq 1$ such that $u_1$ is an \kl{infix} of
    every word $v$ of size at least $k$. In particular, $u_1$ is an \kl{infix}
    of $u_k$, hence the sequence $\seqof{u_i}[i \in \Nat]$ is \kl(sequence){good}.
\end{proof}

\begin{lemma}
    \label{thue-morse:lemma}
    The language $\intro*\LMorse$ of \kl{infixes} of the \kl{Thue-Morse sequence}
    is \kl{downwards closed} for the \kl{infix
    relation}, \kl{well-quasi-ordered} for the \kl{infix relation}, but is not
    \kl(language){bounded}.
\end{lemma}
\begin{proof}
    By construction $\LMorse$ is an \emph{infinite}
    \kl{downwards closed} for the \kl{infix relation}. Let us argue that $\LMorse$ is
    \kl{well-quasi-ordered} for the \kl{infix relation}, but is not \kl(language){bounded}.
    It is \kl{well-quasi-ordered} because it is the set of finite \kl{infixes}
    of a \kl{uniformly recurrent} word, and we can apply \cref{uniformly-recurrent:lem}.

    Assume by contradiction that $\LMorse$ is \kl(language){bounded}. In this case, there exist
    words $w_1, \dots, w_k \in \Sigma^*$ such that $L \subseteq w_1^* \cdots
    w_k^*$. Since $L$ is infinite and \kl{downwards closed}, there exists a
    word $u \in L$ such that $u = w_i^3$ for some $1 \leq i \leq k$. This is absurd
    because $u \infleq \ThueMorse$, which is \kl{cube-free}.
\end{proof}

One may refine our analysis of the \kl{Thue-Morse sequence} to obtain 
precise bounds on the \kl{ordinal invariants} of its language of \kl{infixes}.

\begin{lemma}
    \label{thue-morse-ordinal:lemma}
    The \kl{maximal order type} of $\LMorse$ is $\omegaOrd$,
    the \kl{ordinal height} of $\LMorse$ is $\omegaOrd$,
    the \kl{ordinal width} of $\LMorse$ is $\omegaOrd$.
\end{lemma}
\begin{proof}
    Let us prove that these are upper bounds for the \kl{ordinal invariants} of
    $L$. For the \kl{ordinal height}, it is true for any language $\LMorse$.
    For the \kl{maximal order type}, we remark that
    the maximal length of a \kl{bad sequence} is determined by its first element,
    hence that it is at most $\omegaOrd$.
    Finally, because the \kl{ordinal width} is at most the \kl{maximal order type},
    we conclude that the \kl{ordinal width} is also at most $\omegaOrd$.

    Now, let us prove that these bounds are tight. It is clear that
    $\oHeight{\LMorse} = \omegaOrd$: given any number $n \in \Nat$, one can construct a
    \kl{decreasing sequence} of words in $L$ of length $n$, for instance by
    considering the first $n$ prefixes of the \kl{Thue-Morse sequence} by
    decreasing size.
    Let us now prove that $\oWidth{\LMorse} = \omegaOrd$. Assume by contradiction that
    $\oWidth{\LMorse}$ is finite. Then, $L$ can be written as a finite union of
    \kl{chains} for the \kl{infix relation}, and in particular, $L$ is
    \kl(language){bounded}, which is absurd by \cref{thue-morse:lemma}.
    Finally, because the \kl{ordinal width} is at most the \kl{maximal order
    type}, we conclude that the \kl{maximal order type} of $\LMorse$ is also $\omegaOrd$.
\end{proof}

We prove if the upcoming theorem that the status of the \kl{Thue-Morse
sequence} is actually representative of most \kl{downwards closed} languages
for the \kl{infix relation}. Indeed, we show that the \kl{ordinal invariants}
of such languages are relatively small, and the proof of this fact relies on
the connection between \kl{well-quasi-ordered} languages and \kl{uniformly
recurrent} words.

\begin{theorem}
    \label{small-ordinal-invariants:thm}
    Let $L$ be a \kl{well-quasi-ordered} language for the \kl{infix relation}
    that is \kl{downwards closed}.
    Then, the \kl{maximal order type} of $L$ is strictly less than $\omegaOrd^3$,
    its \kl{ordinal height} is at most $\omegaOrd$,
    and its \kl{ordinal width} is at most $\omegaOrd^2$.
\end{theorem}

\AP A subset $I \subseteq X$ is \intro(subset){directed} if, for every $x,y \in
I$, there exists $z \in I$ such that $x \leq z$ and $y \leq z$. Given a
\kl{well-quasi-order} $(X, \leq)$, one can always decompose $X$ into a finite
union of \intro{order ideals}, that is, non-empty sets $I \subseteq X$ that are
\kl{downwards closed} and \kl(subset){directed} for the relation $\leq$. In our
case, a \kl{well-quasi-ordered} \kl{order ideal} for the \kl{infix relation} is
the set of finite \kl{infixes} of a finite or bi-infinte word $w \in
\Sigma^\Nat \cup \Sigma^\Rel $ (\cref{bi-infinite:lem}). Therefore, we will
first study languages that are described as the set of finite \kl{infixes} of
some (bi-)infinite word (\cref{ultimately-uniformly-recurrent:lem}). To that
end, let us introduce the notation $\intro*\infset{w}$ for the set of finite
\kl{infixes} of a (possibly infinite or bi-infinite) word $w$.

\begin{lemma}
    \label{bi-infinite:lem}
    Let $L \subseteq \Sigma^*$ be an \kl{order ideal} that is \kl{well-quasi-ordered} 
    for the 
    \kl{infix relation}. Then $L$ is the set of finite \kl{infixes}
    of a finite, infinite or bi-infinite word $w$.
\end{lemma}
\begin{proof}
    Let us assume that $L$ is infinite, the case when it is finite 
    is similar, but leads to a finite word.

    Because the alphabet $\Sigma$ is finite, we can enumerate the words of $L$
    as $\seqof{w_i}[i \in \Nat]$. Now, let us build the sequence $\seqof{u_i}[i \in \Nat]$ by induction
    as follows: $u_0 = w_0$, and $u_{i+1}$ is a word that contains $u_i$ and
    $w_i$, which exists in $L$ because $L$ is \kl(subset){directed}. Since $L$ is
    \kl{well-quasi-ordered}, one can extract an infinite set of indices $I
    \subseteq \Nat$ such that $u_i \infleq u_{j}$ for all $i \leq j \in I$.

    We can build a word $w$ as the limit of the sequence $\seqof{u_i}[i \in
    I]$. This word is infinite or bi-infinite, and contains as infixes all the
    words $u_i$ for $i \in I$. Because every word of $L$ is an infix of every
    $u_i$ for a large enough $I$, one concludes that $L$ is contained in the
    set of finite infixes of $w$. Conversely, every finite infix of $w$ is 
    an infix of some $u_i$ by definition of the limit construction, hence
    belongs to $L$ since $u_i \in L$ and $L$ is \kl{downwards closed}.
\end{proof}

\AP We say that an infinite word $w \in \Sigma^\Nat$ is \intro{ultimately
uniformly recurrent} if there exists a bound $N_0 \in \Nat$ such that $w_{\geq
N_0}$ is \kl{uniformly recurrent}.

\begin{lemma}
    \label{ultimately-uniformly-recurrent:lem}
    Let $w \in \Sigma^\Nat$ be an infinite word. 
    Then, the set of finite infixes of $w$ is \kl{well-quasi-ordered} for the \kl{infix relation}
    if and only if $w$ is \kl{ultimately uniformly recurrent}.
\end{lemma}
\begin{proof}

    Assume that $w$ is \kl{ultimately uniformly recurrent}. Consider a sequence
    of words $\seqof{w_i}[i \in \Nat]$ that are finite \kl{infixes} of $w$. Because $w$ is
    \kl{ultimately uniformly recurrent}, there exists a bound $N_0$ such that
    $w_{\geq N_0}$ is \kl{uniformly recurrent}. Let $i < N_0$, we claim that,
    without loss of generality, only finitely many words in the sequence
    $\seqof{w_i}[i \in \Nat]$ can be found starting at the position $i$ in $w$. Indeed, if
    it is not the case, then we have an infinite subsequence of words that are
    all comparable for the \kl{infix relation}, and therefore a \kl{good
    sequence}, because the \kl{infix relation} is \kl{well-founded}. We can
    therefore assume that all words in the sequence $\seqof{w_i}[i \in \Nat]$ are such that
    they start at a position $i \geq N_0$. But then, they are all finite
    \kl{infixes} of $w_{\geq N_0}$, which is a \kl{uniformly recurrent} word,
    whose set of finite \kl{infixes} is \kl{well-quasi-ordered}
    (\cref{uniformly-recurrent:lem}).

    Conversely, assume that the set of finite infixes of $w$ is
    \kl{well-quasi-ordered}. Let us write $\mathsf{Rec}(w)$ the set of finite
    \kl{infixes} of $w$ that appear infinitely often. We can similarly define
    $\mathsf{Rec}(w_{\geq i})$ for any (infinite) suffix of $w$. The sequence
    $R_i \defined \mathsf{Rec}(w_{\geq i})$ is a sequence of \kl{downwards
    closed} sets of finite words, included in the set of finite infixes of $w$
    by definition. Because the latter is \kl{well-quasi-ordered}, there exists
    an $N_0 \in \Nat$, such that $\bigcap_{i \in \Nat} R_i = R_{N_0}$.
    Now, consider $v \defined w_{\geq N_0}$. By construction, 
    every finite infix of $v$ 
    appears infinitely often in $v$. Let us consider 
    some finite infix $u \infleq v$, we claim that there is a bound $N_u$
    on the distance
    between two consecutive occurrences of $u$ in $v$.
    Indeed, if it is not the case, then there exists an infinite sequence
    $\seqof{u x_i u}[i \in \Nat]$ of infixes of $v$, such that $x_i$ is a word that is
    of size $\geq i$.
    Because the finite infixes of $w$ (hence, of $v$) are \kl{well-quasi-ordered},
    one can extract an infinite set of indices $I \subseteq \Nat$
    such that $u x_i u \infleq u x_{j} u$ for all $i \leq j \in I$.
    In particular, $u x_i u \infleq u x_{j} u$ for some $j > |x_i|$, 
    which contradicts the fact that $u x_j u$ coded two consecutive
    occurrences of $u$ in $v$.

    We have shown that for every finite infix $u$ of $v$, there exists a bound
    $N_u$ such that every two occurrences of $u$ in $v$ start at distance at
    most $N_u$. In particular, there exists a bound $M_u$ such that every infix
    of $v$ of size at least $M_u$ contains $u$. We have proven that
    $v$ is \kl{uniformly recurrent}, hence that $w$ is \kl{ultimately uniformly
    recurrent}.
\end{proof}

\begin{lemma}
    \label{small-ordinal-invariants:lem}
    Let $w \in \Sigma^\Nat$ be an \kl{ultimately uniformly recurrent} word.
    Then, the set of finite infixes of $w$ has \kl{ordinal width}
    less than $2 \cdot \omegaOrd$.
\end{lemma}
\begin{proof}
    Let $N_0$ be a bound such that $w_{\geq N_0}$ is \kl{uniformly recurrent}.
    Let us write $\infset{w}$ the set of finite infixes of $w$.
    We prove that $\oWidth{\infset{w}} \leq \omegaOrd + N_0$.
    Let $u_1 \infleq w$ be a finite word. 

    If $u_1$ is an \kl{infix} of $w_{\geq N_0}$, then there exists $k \geq 1$
    such that $u_1$ is an \kl{infix} of every word of size at least $k$. In
    particular, there is finite bound on the length of every sequence of
    incomparable elements starting with $u_1$. We conclude in particular that
    $\infset{w} \setminus \upset{u_1}$ has a finite \kl{ordinal width}.

    Otherwise, $u_1$ can only be found \emph{before} $N_0$. In this case, we
    consider a second element of a \kl{bad sequence} $u_2 \infleq w$, which is
    incomparable with $u_1$ for the \kl{infix relation}. If $u_2$ is an
    \kl{infix} of $w_{\geq N_0}$, then we can conclude as before. Otherwise,
    notice that $u_1$ and $u_2$ cannot start at the same position in $w$
    (because they are incomparable). Continuing this argument, we conclude that
    there are at most $N_0$ elements starting before $N_0$
    at the start of any sequence of
    incomparable elements in $\infset{w}$. We conclude that
    $\oWidth{\infset{w}} \leq \omegaOrd + N_0$.
\end{proof}

Let us briefly discuss the tightness of the bound in
\cref{small-ordinal-invariants:lem}. The \kl{Thue-Morse sequence} over a binary
alphabet $\set{a,b}$ has \kl{ordinal width} $\omegaOrd$. Given a number $N_0
\in \Nat$, one can construct an arbitrarily long \kl{antichain} of words for
the \kl{infix relation} by using a new letter $c$. When concatenating this
(finite) antichain as a prefix of the \kl{Thue-Morse sequence}, one obtains a
new (infinite) word $w$. It is clear that the \kl{ordinal width} of
$\infset{w}$ is now at least $\omegaOrd + N_0$. Hence, the bound $\oWidth{X} <
2 \cdot \omegaOrd$ is tight.

The behaviour of bi-infinite words is very similar to the infinite ones. Given
a bi-infinite word $w \in \Sigma^{\Rel}$, we can consider $w_+ \in \Sigma^\Nat$
and $w_- \in \Sigma^\Nat$ the two infinite words obtained as follows: for all
$i \in \Nat$, $(w_+)_i = w(i)$ and $(w_-)_i = w(-i)$. Note that the two share
the letter at position $0$.

\begin{lemma}
    \label{from-bi-to-single:lem}
    Let $w \in \Sigma^\Rel$ be a bi-infinite word. Then, the set of finite
    infixes of $w$ is \kl{well-quasi-ordered} for the \kl{infix relation} if
    and only if $w_+$ and $w_-$ are two \kl{ultimately uniformly recurrent}
    words. In this case, the \kl{ordinal width}
    of the set of finite infixes of $w$ is less than $3 \cdot \omegaOrd$.
\end{lemma}
\begin{proof}
    Assume that $w_+$ and $w_-$ are \kl{ultimately uniformly recurrent}. Let us
    write $\infset{w}$ the set of finite infixes of $w$. Consider an infinite
    sequence of words $\seqof{u_i}[i \in \Nat]$ in $\infset{w}$. If there is an
    infinite subsequence of words that are all in $\infset{w_+}$, then there
    exists an increasing pair of indices $i < j$ such that $u_i \infleq
    u_j$ because \cref{uniformly-recurrent:lem} applies to $w_+$. Similarly, if
    there is an infinite subsequence of words that are all in $\infset{w_-}$,
    then there exists an increasing pair of indices $i < j$ such that $u_i
    \infleq u_j$ because \cref{uniformly-recurrent:lem} applies to $w_-$ (and
    the \kl{infix relation} is compatible with mirroring). Otherwise, one can
    assume without loss of generality that all words in the sequence have a
    starting position in $w_-$ and an ending position in $w_+$. In this case,
    let us write $(k_i,l_i) \in \Nat^2$ the pair of indices such that $u_i$ is the infix
    of $w$ that starts at position $-k_i$ of $w$ (i.e., $k_i$ of $w_-$) and
    ends at position $l_i$ of $w$ (i.e., $l_i$ of $w_+$).
    Because $\Nat^2$ is a \kl{well-quasi-ordering} with the product ordering,
    there exists $i < j$ such that $k_i \leq k_j$ and $l_i \leq l_j$, 
    in particular, $u_i \infleq u_j$. We have proven that every infinite
    sequence of words in $\infset{w}$ is \kl(wqo){good}, hence $\infset{w}$ is
    \kl{well-quasi-ordered}.

    Conversely, assume that $\infset{w}$ is \kl{well-quasi-ordered}. In
    particular, the subset $\infset{w_+} \subseteq \infset{w}$ is \kl{well-quasi-ordered}.
    Similarly, $\infset{w_-}$ is \kl{well-quasi-ordered} because the \kl{infix
    relation} is compatible with mirroring. Applying
    \cref{ultimately-uniformly-recurrent:lem}, we conclude that both are
    \kl{ultimately uniformly recurrent} words.

    To obtain the upper bound of $3 \cdot \omegaOrd$, we can consider the same
    argument as for \cref{small-ordinal-invariants:lem}. We let $N_0$ be such
    that $w_{\geq N_0}$ and $(w_-)_{\geq N_0}$ are \kl{uniformly recurrent}
    words. In any sequence of incomparable elements of $\infset{w}$, there are
    less than $N_0^2$ elements that are found in $(w_{< N_0})_{> -N_0}$. Then,
    one has to pick a finite \kl{infix} in either $w_{\geq N_0}$ or $w_{\leq
    -N_0}$. Because of \cref{small-ordinal-invariants:lem}, any sequence of
    incomparable elements of these two infinite words has length bounded based
    on the choice of the first element of that sequence. This means that the
    \kl{ordinal width} of $\infset{w}$ is at most $\omegaOrd + \omegaOrd +
    N_0^2$. We conclude that $\oWidth{\infset{w}} < 3 \cdot \omegaOrd$.
\end{proof}

As for \cref{small-ordinal-invariants:lem}, let us briefly discuss
that the bound in \cref{from-bi-to-single:lem} is tight. Indeed, one can
construct a bi-infinite word $w$ by concatenating a reversed \kl{Thue-Morse
sequence} on a binary alphabet $\set{a,b}$, a finite antichain of arbitrarily
large size over a distinct alphabet $\set{c,d}$, and then a \kl{Thue-Morse
sequence} on a binary alphabet $\set{e,f}$. The \kl{ordinal width} of the set
of \kl{infixes} of $w$ is then at least $2 \cdot \omegaOrd + K$, where $K$ is the
size of the chosen antichain.

We are now ready to conclude the proof of \cref{small-ordinal-invariants:thm}.
\begin{proofof}{small-ordinal-invariants:thm}

    It is always true that the \kl{ordinal height} of a language over a finite
    alphabet is at most $\omegaOrd$. Let us now consider a
    \kl{well-quasi-ordered} language $L$ that is \kl{downwards closed} for the
    \kl{infix relation}. Because it is a \kl{well-quasi-ordered} set, it can be
    written as a finite union of \kl{order ideals} $L = \bigcup_{i = 1}^n L_i$.

    For every such \kl(order){ideal} $L_i$, we can apply
    \cref{bi-infinite:lem}, and conclude that $L_i$ is the set of finite
    \kl{infixes} of a finite, infinite or bi-infinite word $w_i$. We can then
    directly conclude that $\oWidth{L_i}$ less than $\omegaOrd$ (in the case of
    a finite word), less than $2 \cdot \omegaOrd$ (in the case of an infinite
    word thanks to \cref{small-ordinal-invariants:lem}), or less than $3 \cdot
    \omegaOrd$ (in the case of a bi-infinite word, thanks to
    \cref{from-bi-to-single:lem}). In any case,
    we have the bound $\oWidth{L_i} < 3 \cdot \omegaOrd$.

    Now, $\oWidth{L} \leq \sum_{i = 1}^n \oWidth{L_i} < 3 \cdot \omegaOrd <
    \omegaOrd^2$. We conclude that the \kl{ordinal width} of $L$ is less than
    $\omegaOrd^2$. Finally, the inequality $\oType{L} \leq \oWidth{L} \oComProd
    \oHeight{L} < \omega \oComProd \omega^2 = \omegaOrd^3$ allows us to conclude.
\end{proofof}

\AP As we have demonstrated, infinite (or bi-infinite words) can be used to
represent languages that are \kl{well-quasi-ordered} for the \kl{infix
relation} by considering their set of finite \kl{infixes}. We can leverage the
theory of \kl{automatic sequences}, that is tailored to study the regularity of
infinite words to build a decision procedure for \kl{well-quasi-ordered}
languages for the \kl{infix relation}. A sequence $w \in \Sigma^\Nat$ is
\intro(sequence){automatic} whenever there exists a base $b \in \Nat$ and a
finite automaton that can compute the $i$-th letter of $w$ given as input the
number $i$ in base $b$. The \kl{Thue-Morse sequence} is an example of an
\kl{automatic sequence}.

\begin{theorem}
    \label{automatic-wqo:thm}
    Given an \kl{automatic sequence} $w \in \Sigma^{\Nat}$, one can decide
    whether the set of \kl{infixes} of $w$ is \kl{well-quasi-ordered} for the
    \kl{infix relation}.
\end{theorem}
\begin{proof}
    Because of \cref{ultimately-uniformly-recurrent:lem}, it suffices
    to decide whether $w$ is \kl{ultimately uniformly recurrent}.
    We can rewrite this as a question on the \kl{automatic sequence} $w$
    as follows:
    \begin{align*}
        &\exists N_0,                   &   \text{ultimately} \\
        &\forall i_s \geq N_0,          &   \text{for every infix (start) } u \\
        &\forall i_e > i_s,             &   \text{for every infix (end) }   u \\
        &\exists k \geq 1,              &   \text{there exists a bound} \\
        &\forall j_s \geq N_0,          &   \text{for every other infix (start) } v \\
        &\forall j_e \geq j_s + k,      &   \text{of size at least $k$} \\
        &\exists l \geq 0,              &   \text{there exists a position in } v \\
        &\forall 0 \leq m < i_e - i_s,  &   \text{where } u \text{ can be found} \\
        &j_s + m + l < j_e \land
        w(i_s + m) = w(j_s + m + l) \quad .
    \end{align*}
    Because $w$ is computable by a finite automaton, one can reduce the above
    formula to a regular language, for which it suffices to check emptiness, which
    is decidable.
\end{proof}

Note that being \kl{uniformly recurrent} still leaves open a lot of potential
behaviours, for instance, Sturmian words are \kl{uniformly recurrent}, but are
not \kl(sequence){automatic}. Interestingly enough, the \kl{ordinal invariants}
of the \kl{automatic sequences} are not smaller than the ones of \kl{ultimately
uniformly recurrent} words: this contrasts with the case of languages that are
\kl{well-quasi-ordered} for the \kl{infix relation} in general, where the
hypothesis of being representable (by a \kl{regular language}, or an
\kl{amalgamation system}) will drastically reduce the \kl{ordinal invariants}
of said language (\cref{ordinal-invariants-bounded:cor}, \cref{infix-amalgamation:thm}).
\section{Infixes and Amalgamation Systems}
\label{infixes-amalgamation:sec}

\AP In this section, we are going to design an effective decision procedure for
\kl{well-quasi-ordering} by the \kl{infix relation}. To that end, the first
requirement is to fix a way to represent languages $L \subseteq \Sigma^*$.
Traditionally, one would use finite automata, context-free grammars, or for the
more adventurous, vector addition systems with states. However, our proof
technique will only require us to have a way to ``glue'' together runs of the
system to ``pump'' them and produce new runs: this is the usual pumping lemma
in automata theory, and Ogden's lemma for context-free grammars \cite{OGDEN68}.
It turns out that there is a rather large family of systems for which pumping
arguments based on so-called \emph{minimal runs} exist: they are called
\emph{amalgamation systems} and were recently introduced by Anand, Schmitz,
Schütze, and Zetzsche \cite{ASZZ24}.

\AP Our first result, of theoretical nature, is that \kl{amalgamation systems}
cannot define \kl{well-quasi-ordered} languages that are not
\kl(language){bounded}. This implies that all the results of
\cref{infixes-bounded:sec}, and in
particular \cref{bounded-language:thm}, can safely be applied to
\kl{amalgamation systems}.

\begin{theorem}[label=infix-amalgamation:thm,restate=infix-amalgamation:thm]
    \proofref{infix-amalgamation:thm}
    Let $L \subseteq \Sigma^*$ be a language recognized by an 
    \kl{amalgamation system}.
    If $L$ is \kl{well-quasi-ordered} by the \kl{infix relation} then $L$ is
    \kl(language){bounded}.
\end{theorem}

\AP Our second focus is of practical nature: we want to give a decision
procedure for being \kl{well-quasi-ordered}. This will require us to introduce
\emph{effectiveness assumptions} on the \kl{amalgamation systems}. While most
of them will be innocuous, an important consequence is that we have to consider
\emph{classes of languages} rather than individual ones, for instance: the
class of all regular language, or the class of all context-free languages. Such
classes will be called \kl{effective amalgamative classes} (\kcref{effective
amalgamative classes}). In the following theorem, we prove that under such
assumptions, testing \kl{well-quasi-ordering} is inter-reducible to testing
whether a language of the class is empty, which is usually the simplest
problem for a computational model.

\begin{theorem}[restate=infix-wqo-is-emptiness:thm,label=infix-wqo-is-emptiness:thm]
	Let $\mathcal{C}$ be an \kl{effective amalgamative class} of languages.
    Then the following are equivalent:
	\begin{enumerate}
        \item\label{wqo-infix-decidable} \kl[wqo]{Well-quasi-orderedness} of the \kl{infix relation} is decidable for languages in $\mathcal{C}$.
        \item\label{wqo-prefix-decidable} \kl[wqo]{Well-quasi-orderedness} of the \kl{prefix relation} is decidable for languages in $\mathcal{C}$.
        \item\label{emptiness-decidable} Emptiness is decidable for languages in $\mathcal{C}$.
	\end{enumerate}
\end{theorem}

\AP We say that a \kl{strongly effective amalgamative class} is an
\kl{effective amalgamative class} for which the emptiness problem is decidable.
Let us immediately remark that the class of \kl{regular languages} is
\kl(amalg){strongly effective}, and so is the class of context-free languages.
Therefore, \cref{infix-wqo-is-emptiness:thm} provides a concrete decision
procedure for these classes.

\subsection{Amalgamation Systems}
\label{amalgamation-systems:subsec}

Let us now formally introduce the notion of \kl{amalgamation systems}, and
recall some results from \cite{ASZZ24} that will be useful for the proof of
\cref{infix-amalgamation:thm}. The notion of \kl{amalgamation system} is
tailored to produce \emph{pumping arguments}, which is exactly what our
\cref{pumping-periods:lem} talks about. At the core of a pumping argument,
transitions taken in a finite state automaton. Continuing on the analogy with
finite automata, there is a natural ordering between runs, i.e., a run is
smaller than another one if one can ``delete'' loops of the larger run to obtain
the other. Typical pumping arguments then rely on the fact that
\emph{minimal} runs are of finite size, and that all other runs are
obtained by ``gluing'' loops to minimal runs. Generalizing this notion yields the 
notion of \kl{amalgamation systems}.

\AP Let us recall that over an alphabet $(\Sigma, =)$ a \kl{subword embedding}
between two words $u \in \Sigma^*$ and $v \in \Sigma^*$ is a function $\rho
\colon \range{\card{u}} \to \range{\card{v}}$ such that $u_i = v_{\rho(i)}$ for
all $i \in \range{\card{u}}$. We write $\intro*\HigEmb(u,v)$ the set of all
\kl{subword embeddings} between $u$ and $v$. It may be useful to notice that
the set of finite words over $\Sigma$ forms a category when we consider
\kl{subword embeddings} as morphisms, which is a fancy way to state that
$\mathrm{id} \in \HigEmb(u,u)$ and that $f \circ g \in \HigEmb(u,w)$ whenever
$g \in \HigEmb(u,v)$ and $f \in \HigEmb(v,w)$, for any choice of words
$u,v,w \in \Sigma^*$.

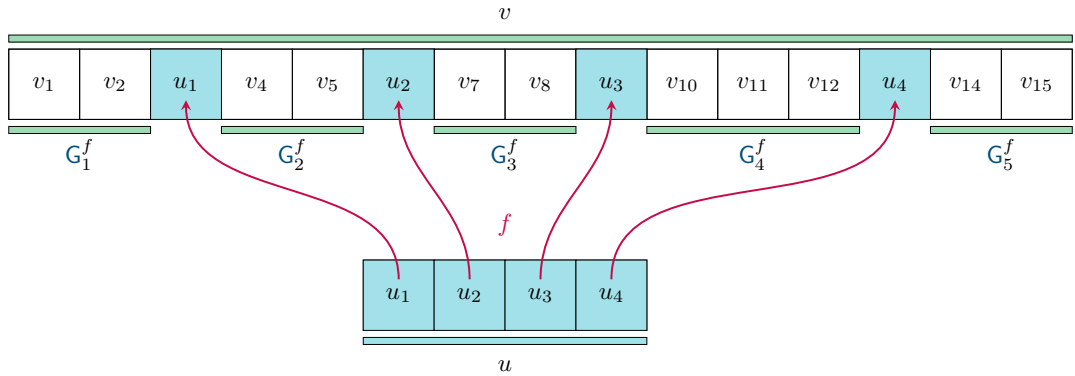
\begin{figure}
    \centering
    \begin{tikzpicture}[
        embedding/.style={
            thick, ->, >=stealth,
            A2
        },
        gapWord/.style={
            fill=B5bg,
        },
        startWord/.style={
            fill=D4bg,
        },
        ]

    \node (u) at (2,-0.5) {$u$};
    \draw[startWord] (0,-0.1) rectangle (4,-0.2);
    \draw (0,0) rectangle (4,1);
    \foreach[count=\xi] \x in {0,1,2,3} {
        \pgfmathsetmacro{\xp}{int(\x + 1)}
        \draw[startWord] (\x,0) rectangle (\xp,1);
        \node (u\xi) at (\x+0.5,0.5) {$u_{\xi}$};
    }

    \begin{scope}[yshift=3cm,xshift=-5cm]
        \draw (0,0) rectangle (15,1);
        \foreach[count=\xi] \x in {0,1,2,3,4,5,6,7,8,9,10,11,12,13,14} {
            \pgfmathsetmacro{\xp}{int(\x + 1)}
            \draw (\x,0) rectangle (\xp,1);
            \node (v\xi) at (\x+0.5,0.5) {$v_{\xi}$};
        }
    \end{scope}

    \foreach \xi/\yi in {1/3,2/6,3/9,4/13} {
        \draw[startWord] ({\yi - 5},3) rectangle ({\yi - 6},4);
        \node (uv\xi) at ({\yi - 5.5},3.5) {$u_{\xi}$};
        \draw[embedding] (u\xi) to[in=-90,out=90] (v\yi);
    }

    \foreach[count=\count] \xi/\size in {1/2,4/2,7/2,10/3,14/2} {
        \draw[gapWord] ({\xi - 6 + \size},2.9) rectangle ({\xi - 6},2.8);
        \node (G\xi) at ({\xi - 6 + 0.5*\size},2.5) {$\GapWord{f}{\count}$};
    }

    \draw[gapWord] (-5,4.1) rectangle (10,4.2);
    \node (v) at (2,4.5) {$v$};

    \node[embedding] (f) at (2,1.5) {$f$};

\end{tikzpicture}
    \caption{The \kl{gap words} resulting from a \kl{subword embedding} between two 
    finite words.}
    \label{gap-word-embedding:fig}
\end{figure}

\AP Given a \kl{subword embedding} $f \colon u \to v$ between two words $u$ and
$v$, there exists a unique decomposition $v = \GapWord{f}{0} u_1 \GapWord{f}{1}
\cdots \GapWord{f}{k-1} u_k \GapWord{f}{k}$ where $\GapWord{f}{i} =
v_{f(i)+1} \cdots v_{f(i+1)-1}$ for all $1 \leq i \leq k-1$, $\GapWord{f}{k} =
v_{[f(k)+1} \cdots v_{\card v}$, and $\GapWord{f}{0}   = v_1 \cdots v_{f(1)-1}$. We say that
$\intro*\GapWord{f}{i}$ is the $i$-th \intro{gap word} of $f$. We encourage the
reader to look at \cref{gap-word-embedding:fig} to see an example of the
\kl{gap words} resulting from a \kl{subword embedding} between two words. These
\kl{gap words} will be useful to describe how and where runs of a system
(described by words) can be combined.

\begin{figure}
    \centering
    \begin{tikzpicture}[
        embedding/.style={
            thick, ->, >=stealth,
            A2
        },
        gapWord/.style={
            fill=B5bg,
        },
        startWord/.style={
            fill=D4bg,
        },
        embF/.style={
            A1
        },
        embG/.style={
            A2
        },
        gapF/.style={
            fill=A1bg,
        },
        gapG/.style={
            fill=A2bg,
        },
        ]

    \node (u) at (2,-0.5) {$u$};
    \draw[startWord] (0,-0.1) rectangle (4,-0.2);
    \draw (0,0) rectangle (4,1);
    \foreach[count=\xi] \x in {0,1,2,3} {
        \pgfmathsetmacro{\xp}{int(\x + 1)}
        \draw[startWord] (\x,0) rectangle (\xp,1);
        \node (u\xi) at (\x+0.5,0.5) {$u_{\xi}$};
    }

    \begin{scope}[yshift=6cm,xshift=-15cm]
        \draw (0,0) rectangle (15,1);
        \foreach[count=\xi] \x in {0,1,2,3,4,5,6,7,8,9,10,11,12,13,14} {
            \pgfmathsetmacro{\xp}{int(\x + 1)}
            \draw (\x,0) rectangle (\xp,1);
            \node (v\xi) at (\x+0.5,0.5) {$v_{\xi}$};
        }
        \foreach \xi/\yi in {1/3,2/6,3/9,4/13} {
            \draw[startWord] (\yi,0) rectangle ({\yi - 1},1);
            \node (uv\xi) at ({\yi - 0.5},0.5) {$u_{\xi}$};
            \draw[embedding,embF] (u\xi) to[in=-90,out=90] (v\yi);
        }
        \foreach[count=\count] \xi/\size in {1/2,4/2,7/2,10/3,14/2} {
            \draw[gapWord,gapF] ({\xi - 1 + \size},-0.1) rectangle ({\xi - 1},-0.2);
            \node (G\xi) at ({\xi - 1 + 0.5*\size},-0.5) {$\GapWord{f}{\count}$};
        }

        \node (v) at (-0.5,0.5) {$v$};

    \end{scope}

    \node[embedding,embF] (f) at (0,1.5) {$f$};
    \node[embedding,embG] (g) at (4,1.5) {$g$};
    \node[embedding,embG] (h) at (18,10) {$h$};
    \node[embedding,embF] (k) at (-13,10) {$k$};

    \begin{scope}[yshift=6cm,xshift=5cm]
        \draw (0,0) rectangle (15,1);
        \foreach[count=\xi] \x in {0,1,2,3,4,5,6,7,8,9,10,11,12,13,14} {
            \pgfmathsetmacro{\xp}{int(\x + 1)}
            \draw (\x,0) rectangle (\xp,1);
            \node (w\xi) at (\x+0.5,0.5) {$w_{\xi}$};
        }
        \foreach \xi/\yi in {1/3,2/6,3/9,4/13} {
            \draw[startWord] (\yi,0) rectangle ({\yi - 1},1);
            \node (uw\xi) at ({\yi - 0.5},0.5) {$u_{\xi}$};
            \draw[embedding,embG] (u\xi) to[in=-90,out=90] (w\yi);
        }
        \foreach[count=\count] \xi/\size in {1/2,4/2,7/2,10/3,14/2} {
            \draw[gapWord,gapG] ({\xi - 1 + \size},-0.1) rectangle ({\xi - 1},-0.2);
            \node (G\xi) at ({\xi - 1 + 0.5*\size},-0.5) {$\GapWord{g}{\count}$};
        }
        \node (w) at (15.5,0.5) {$w$};
    \end{scope}

    \begin{scope}[yshift=13cm,xshift=5cm]
        \draw (-15,0) rectangle (11,1);
        \foreach[count=\xi] \x in {-15,...,10} {
            \pgfmathsetmacro{\xp}{int(\x + 1)}
            \draw (\x,0) rectangle (\xp,1);
            \node (z\xi) at (\x+0.5,0.5) {$z_{\xi}$};
        }
        \foreach[count=\count] \xi in {5,10,15,22} {
            \draw[startWord] ({\xi - 15},1) rectangle ({\xi - 16},0);
            \node (vwz\xi) at ({\xi - 15.5},0.5) {$u_{\count}$};
        }
        \foreach[count=\count] \xi/\size in {3/2,6/2,11/2,19/3,23/2} {
            \draw[gapWord,gapG] ({\xi - 16 + \size},1.1) rectangle ({\xi - 16},1.2);
            \node (G\xi) at ({\xi - 16 + 0.5*\size},1.5) {$\GapWord{g}{\count}$};
        }
        \foreach[count=\count] \xi/\size in {1/2,8/2,13/2,16/3,25/2} {
            \draw[gapWord,gapF] ({\xi - 16 + \size},1.1) rectangle ({\xi - 16},1.2);
            \node (G\xi) at ({\xi - 16 + 0.5*\size},1.5) {$\GapWord{f}{\count}$};
        }
    \end{scope}

    \foreach \xi/\yi in 
    {1/1,2/2,3/5,4/8,5/9,6/10,7/13,8/14,9/15,10/16,11/17,12/18,13/22,14/25,15/26} {
        \draw[embedding,embF] (v\xi) to[in=-90,out=90] (z\yi);
    }
    \foreach \xi/\yi in 
    {1/3,2/4,3/5,4/6,5/7,6/10,7/11,8/12,9/15,10/19,11/20,12/21,13/22,14/23,15/24} {
        \draw[embedding,embG] (w\xi) to[in=-90,out=90] (z\yi);
    }

\end{tikzpicture}
    \caption{We illustrate how 
        embeddings $f$ and $g$ between runs of an
        \kl{amalgamation system} can be glued
        together, seen on their canonical decomposition.
    }
    \label{amalgamation-runs:fig}
\end{figure}
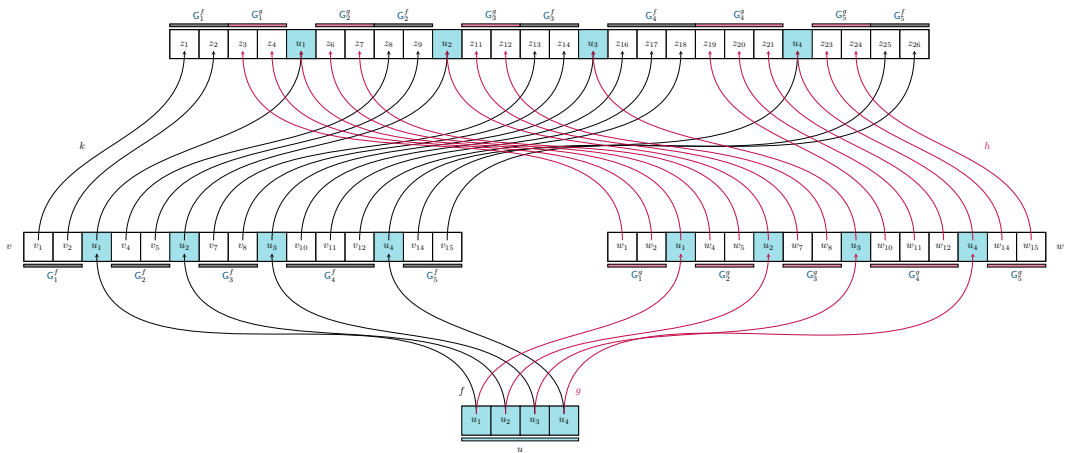

\begin{definition}
    An \intro{amalgamation system}
    is a tuple $(\Sigma, R, \canrun, E)$ where
    $\Sigma$ is a finite alphabet,
    $R$ is a set of so-called \emph{runs},
    $\canrun \colon R \to (\Sigma \uplus \set{\cansep})^*$ is a 
    function computing a \intro{canonical decomposition} of a run,
    and $E$ describes the so-called \intro{admissible embeddings} between runs: If $\rho$ and $\sigma$ are runs from $R$, then $E(\rho, \sigma)$ is a subset of the subword embeddings between $\canrun[\rho]$ and $\canrun[\sigma]$. We write $\rho \intro*\runleq \sigma$ if $E(\rho, \sigma)$ is non-empty. If we want to refer to a specific embedding $f \in E(\rho, \sigma)$, we also write $\rho \runleq_f \sigma$.
    Given a run $r \in R$, and $i \in \range[0]{\card{\canrun(r)}}$, 
    the \intro{gap language} of $r$ at position $i$ is $\intro*\GapLanguage{r}{i} \defined
    \setof{\GapWord{f}{i}}{\exists s \in R. \exists f \in E(r,s) }$.
    An \kl{amalgamation system} furthermore satisfies the following 
    properties:
    \begin{enumerate}
        \item \emph{$(R, E)$ Forms a Category.}
            For all $\rho, \sigma, \tau \in R$,
            $\mathrm{id} \in E(\rho,\rho)$,
            and whenever $f \in E(\rho,\sigma)$ and $g \in E(\sigma,\tau)$,
            then $g \circ f \in E(\rho,\tau)$.
        \item \emph{Well-Quasi-Ordered System.}
            $(R, \runleq)$ is a well-quasi-ordered set.
        \item \emph{Concatenative Amalgamation.}
            Let $\rho_0, \rho_1, \rho_2$ be runs 
            with $\rho_0 \runleq_f \rho_1$ 
            and $\rho_0 \runleq_g \rho_2$.
            Then for all $0 \leq i \leq \card{\canrun(\rho_0)}$,
            there exists a run $\rho_3 \in R$ 
            and embeddings $\rho_1 \runleq_{g'} \rho_3$
            and $\rho_2 \runleq_{f'} \rho_3$ 
            satisfying two conditions:
            (a) $g' \circ f = f' \circ g$ (we write $h$ for this composition) and
            (b) for every $0 \leq j \leq \card{\canrun[\rho_0]}$, 
            the gap word $\GapWord{h}{j}$
            is either $\GapWord{f}{j} \GapWord{g}{j}$
            or $\GapWord{h}{j} = \GapWord{g}{j} \GapWord{f}{j}$. 
            Specifically, for $i$ we may fix $\GapWord{h}{i} = \GapWord{f}{i} \GapWord{g}{i}$.
            We refer to \cref{amalgamation-runs:fig} for an illustration 
            of this property.
    \end{enumerate}

	The \emph{yield} of a run is obtained by projecting away the separator symbol \cansep~from the canonical decomposition, i.e. $\intro*\yieldrun(\rho) = \project_\Sigma( \canrun[\rho])$. The language recognized by an \kl{amalgamation system} is $\yieldrun(R)$.
    
    We say a language $L$ is an \intro{amalgamation language} 
    if there exists an \kl{amalgamation system} recognizing it. 
\end{definition}

Intuitively, the definition of an amalgamation system allows the comparison of
runs, and the proper ``gluing'' of runs together to obtain new runs. A number of well-known language classes can be seen to be recognized by \kl{amalgamation
systems}, e.g., regular languages \cite[Theorem 5.3]{ASZZ24}, reachability and coverability languages of VASS \cite[Theorem 5.5]{ASZZ24}, and context-free languages \cite[Theorem 5.10]{ASZZ24}. For this paper to be self-contained, we will also
recall how runs of a finite state automaton can be understood as an
\kl{amalgamation system}.

\begin{example}[{\cite[Section 3.2]{ASZZ24}}]
    Let $A = (Q, \delta, q_0, F)$ be a finite state automaton over a finite
    alphabet $\Sigma$. Let $\Delta$ be the set of transitions $(q_1, a, q_2)
    \in Q \times \Sigma \times Q$,
    and $R \subseteq \Delta^*$ be the set of 
    words over transitions that start with the initial state $q_0$,
    end in a final state $q_f \in F$, and such that the end state of a
    letter is the start state of the following one.
    The canonical decomposition $\canrun$
    is defined as a morphism from $\Delta^*$ to $\Sigma^*$
    that maps $(q,a,p)$ to $a$. 
    Because of the one-to-one correspondence of steps of a run $\rho$ and letters in its \kl{canonical decomposition}, 
    we may treat the two interchangeably.
    Finally, given two runs $\rho$ and $\sigma$ of the automaton,
    we say that an embedding $f \in \HigEmb(\canrun(\rho), \canrun(\sigma))$
    belongs to $E(\rho,\sigma)$ when
    $f$ is also defining an embedding from $\rho$ to $\sigma$ as words in $\Delta^*$.

    The system $(\Sigma, R, E, \canrun)$ is an \kl{amalgamation system},
    whose language is precisely the language of words recognized
    by the automaton $A$.
\end{example}
\begin{proof}
    By definition, the embeddings inside $E(\rho,\sigma)$ are in
    of $\HigEmb(\canrun(\rho), \canrun(\sigma))$, and they compose properly.
    Because $\Delta = Q \times \Sigma \times Q$ is finite, it is 
    a \kl{well-quasi-ordering} when equipped with the equality relation, and 
    we conclude that $\Delta^*$ with $\higleq$ is a \kl{well-quasi-order}
    according to Higman’s Lemma \cite{HIG52}.
    
    Let us now move to proving that the system satisfies the amalgamation
    property. Given three runs $\rho,\sigma,\tau \in R$, and two embeddings $f \in E(\rho,\sigma)$
    and $g \in E(\rho,\tau)$, we want to construct an amalgamated run $\sigma \vee \tau$.
    Because letters in the run $\rho$ respect the transitions of the automaton
    (i.e., if the letter $i$ ends in state $q$, then the letter $i+1$ starts in
    state $q$), then the \kl{gap word} at position $i$ starts in state $q$ and
    ends in state $q$ too. This means that for both embeddings
    $f$ and $g$, the \kl{gap words} are read by the automaton by looping
    on a state. In particular, these loops can be taken in any order
    and continue to represent a valid run. That is, we can even select
    the order of concatenation in the amalgamation for \emph{all} 
    $0 \leq i \leq \card{\canrun(\rho)}$ and not just for one separately.

    We conclude by remarking that 
    the language of this amalgamation system is
    the set of $\yieldrun(R)$, 
    because $R$ is the set of valid runs of the automaton,
    and $\yieldrun(\rho)$ is the word read along a run $\rho$.
\end{proof}

\subsection{$\infleq$-Well-Quasi-Ordered Amalgamation Systems}

We can now show a simple lemma that illuminates much of the structure of amalgamation systems whose language is well-quasi-ordered by $\infleq$.

\begin{lemma}
	\label{gap-words-prefix-ordered:lem}
	Let $L$ by an \kl{amalgamation language} recognized by $(\Sigma, R, E, \canrun)$ that is well-quasi-ordered by $\infleq$. Let $\rho$ be a run with $\canrun[\rho] = a_1 \cdots a_n$, and let $\sigma, \tau$ be runs with $\rho \runleq_f \sigma$ and $\rho \runleq_g \sigma$. 
	
	For any $0 \leq \ell \leq n$, we have $\GapWord{f}{\ell} \infleq \GapWord{g}{\ell}$ or vice versa.
\end{lemma}

\begin{proof}
	Write $u$ for $\GapWord{f}{\ell}$ and $v$ for $\GapWord{g}{\ell}$. 
	We may assume that both $u$ and $v$ are non-empty, as otherwise the lemma holds trivially.
	Then, for all $k \in \Nat$, there exists a run with canonical decomposition
	$$
	w_k = L_0 a_1 \cdots a_n L_n,
	$$
	where $L_i \in \set{vv u^k, vu^kv, u^k vv}$ and specifically $L_\ell = vu^kv$.
	
	From \cref{pumping-periods:lem}, we may conclude that there are a finite number of words $x, y,$ and $w$ 
	such that each $w_k$ is contained in a language 
	$\InfPeriodChain{x}w\InfPeriodChain{y}$.
	
	As there is an infinite number of words $w_k$, 
	we may fix $x, y,$ and $w$ and an infinite subset $I \subseteq \Nat$ 
	such that $\set{w_i \mid i \in I} \subseteq \InfPeriodChain{x}w\InfPeriodChain{y}$. 
	This implies that either for infinitely many $m \in \Nat$, $u^m v \in \InfPeriodChain{y}$ 
	or for infinitely many $m$, $v u^m \in \InfPeriodChain{x}$. 
	
	In either case, we may conclude that either $u \infleq v$ or $v \infleq u$: Let $m, n \in \Nat$
	such that $m < n$ and $u^m v, u^n v \in \InfPeriodChain{y}$ (the case for $v u^m$ and $v u^n$ 
	proceeding analogously). Without loss of generality, assume that $\card{u^m}$ and $\card{u^n}$ are
	multiples of $\card{y}$. We therefore find $p \prefleq y, s \suffleq y$ such that $u^m, u^n \in sy^*p$, 
	ergo $ps = y$.
	In other words, we can write $u^m = (sp)^{m'}, u^n = (sp)^{n'}$. As $u^mv \in \InfPeriodChain{y}$, it 
	follows that $v$ is a prefix of some word in $(sp)^*$. Hence either $v$ is a prefix of $u$ or $u$ vice versa.
\end{proof}

\begin{proofof}{infix-amalgamation:thm}
    Assume that $L$ is \kl{well-quasi-ordered} by the \kl{infix relation},
    and obtained by an \kl{amalgamation system}
    $(\Sigma, R, E, \canrun)$.

    Let us consider the set $M$ of minimal runs for the relation $\leq_E$,
    which is finite because the latter is a \kl{well-quasi-ordering}. 
    By \cref{gap-words-prefix-ordered:lem}, we know that for each minimal run $\rho \in M$,
    each gap language $\GapLanguage{\rho}{i}$ of $\rho$ is totally ordered by $\infleq$.
    Adapting the proof of language boundedness from \cite[Section 4.2]{ASZZ24}, we may conclude that $\GapLanguage{\rho}{i} \subseteq \InfPeriodChain{w}$ for some $w \in \GapLanguage{\rho}{i}$.
    As $\InfPeriodChain{w}$ is language bounded and this property is stable under subsets, concatenation and finite union,
    we can conclude that $L$ is bounded as well.
\end{proofof}

If we additionally assume that such a language is closed under infixes, we obtain an even stronger structure: All such languages are regular:

\begin{lemma}
    \label{dwclosed-infixes-wqo:lem}
    Let $L \subseteq \Sigma^*$ be a \kl{downwards closed} language for the
    \kl{infix relation} that is \kl{well-quasi-ordered}. Then, the following
    are equivalent:
    {\renewcommand{\theenumi}{\roman{enumi}}
     \renewcommand{\labelenumi}{(\theenumi)}
    \begin{enumerate}
        \item\label{dwci-reg:item} $L$ is a \kl{regular language},
        \item\label{dwci-aml:item} $L$ is recognized by \emph{some} \kl{amalgamation system},
        \item\label{dwci-bod:item} $L$ is a \kl{bounded language},
        \item\label{dwci-uoc:item} There exists 
            a finite set $E \subseteq (\Sigma^*)^3$
            such that $L = \bigcup_{(x,u,y) \in E} \InfPeriodChain{x} u \InfPeriodChain{y}$.
    \end{enumerate}
    }
\end{lemma}
\begin{proof}
    It is clear that \cref{dwci-reg:item} $\Rightarrow$ \cref{dwci-aml:item}
    because regular languages are recognized by finite automata, and finite
    automata are a particular case of \kl{amalgamation systems}.
    The implication \cref{dwci-aml:item} $\Rightarrow$ \cref{dwci-bod:item}
    is the content of \cref{infix-amalgamation:thm}.
    The implication \cref{dwci-bod:item} $\Rightarrow$ \cref{dwci-uoc:item}
    is \cref{bounded-language:lem}.
    Finally, the implication \cref{dwci-uoc:item} $\Rightarrow$ \cref{dwci-reg:item}
    is simply because a \kl{downwards closed} language 
    that is a finite union of products of \kl{chains} is a regular language.

    Indeed, assume that
    $L$ is \kl{downwards closed} and included in a finite union of sets of the form
    $\InfPeriodChain{x} u \InfPeriodChain{y}$ where $x,y,u$ are possibly empty words.
    We can assume without loss of generality that
    for every $n$, $x^n u y^n$ is in $L$, otherwise, we have a bound on the maximal $n$ such that
    $x^n u y^n$ is in $L$, and we can increase the number of languages in the union, replacing $x$ or $y$
    with the empty word as necessary.
    Let us write $L' \defined \bigcup_{i = 1}^k x_i^* u_i y_i^*$. Then, $L'
    \subseteq L$ by construction. Furthermore, $L \subseteq \dwset{L'}$, also
    by construction. Finally, we conclude that $L = \dwset{L'}$ because $L$ is
    \kl{downwards closed}. Now, because $L'$ is a \kl{regular language}, and 
    \kl{regular languages} are closed under \kl{downwards closure}, we conclude
    that $L$ is a \kl{regular language}.
\end{proof}

\AP Combining
\cref{thue-morse:lemma,dwclosed-infixes-wqo:lem}, we can
conclude that the collection of \kl{infixes} of the \kl{Thue-Morse sequence}
cannot be recognized by \emph{any} \kl{amalgamation system}.

\subsection{Effective Decision Procedures}
\label{infixes-amalgamation-effective:subsec}

\AP Let us now introduce our effectiveness assumptions on \kl{amalgamation
systems}. We follow the approach of \cite{ASZZ24} and require that an
\kl{amalgamation system} $(\Sigma, R, E, \canrun)$ is effective when $R$ is
recursively enumerable, the function $\canrun(\cdot)$ is computable, and for
any two runs $\rho, \sigma \in R$, the set $E(\rho,\sigma)$ is computable.

\AP We say that a class $\mathcal{C}$ of languages is an \intro{effective
amalgamative class} whenever for every $L \in \mathcal{C}$, there exists an
effective \kl{amalgamation system} recognizing $L$, and such that $\mathcal{C}$
is \kl{effectively closed under rational transductions}. Recall that a
\intro{rational transduction} is a \emph{relation} $R \subseteq \Sigma^* \times
\Gamma^*$ that can be defined by a (nondeterministic) finite automaton with
output \cite[Chapter 5, page 64]{BERST79}. A class of languages $\mathcal{C}$
is \intro{effectively closed under rational transductions} when, for every
language $L \in \mathcal{C}$, and every rational transduction $R \subseteq
\Sigma^* \times \Gamma^*$, the image of $L$ through $R$ --- that is, $\setof{v
\in \Gamma^*}{\exists u \in L. (u,v) \in R}$ --- is in $\mathcal{C}$ and
effectively computable. In particular, it implies that the class contains
languages over arbitrary (finite) alphabets.

\AP A class $\mathcal{C}$ of languages is \intro(amalg){strongly effective}
whenever the emptiness problem for languages in $\mathcal{C}$ is decidable.
This notion is interesting because usual language problems such as boundedness
or simultaneous boundedness are decidable for \kl{strongly effective
amalgamative classes}~\cite{ASZZ24}. 

\begin{proofof}{infix-wqo-is-emptiness:thm}
	\cref{emptiness-decidable} We first show $\Rightarrow$ \cref{wqo-infix-decidable}. We aim to make the inclusion test of \cref{bounded-language:eq} of \cref{bounded-language:thm} effective. 
	Let $R(n,m,N_0) \defined \bigcup_{x,y \in \Sigma^{\leq n}} \bigcup_{u \in
    \Sigma^{\leq m \times N_0}} \InfPeriodChain{x} u \InfPeriodChain{y} \cup
    \InfPeriodChain{x}u \cup u\InfPeriodChain{x}$. For any concrete values of
    the bounds $n$, $m$, and $N_0$, this language is regular. The map $L
    \mapsto L \cap \Sigma^* \setminus R(n,m,N_0)$  is a \kl{rational
    transduction} because $\Sigma^* \setminus R(n,m,N_0)$ is regular. Since
    $\mathcal{C}$ is \kl{closed under rational transductions}, we can therefore
    reduce the inclusion to emptiness of this language. However, we need to
    find these bounds first.
	
    To determine values for $n$ and $m$, we first test if $L$ is
    \kl(language){bounded}. Since emptiness is decidable, we can apply the
    algorithm in~\cite[Section 4.2]{ASZZ24} to decide if $L$ is
    \kl(language){bounded}. If $L$ is \kl(language){bounded}, this algorithm
    yields words $w_1, \ldots w_n$ such that $L \subseteq w_1^* \cdots w_n^*$
    and therefore yields also the bounds in questions: $n$ is the number of
    words, and $m$ is the maximal length of a word $w_i$ where $1 \leq i \leq
    n$. If $L$ is not bounded, then $L$ cannot be \kl{well-quasi-ordered} by
    the \kl{infix relation} because of \cref{infix-amalgamation:thm} and we
    immediately return false.
	
    To determine the value for $N_0$, we then compute the downward closure (with respect to subwords) of $L$. 
    This is effective and yields a finite-state automaton. 
    Recall that $N_0$ is the maximum number of repetitions of a word $w_i$ that can not be iterated arbitrarily often. 
    This value is therefore bounded above by the length of the longest simple path in this automaton.
    
    \cref{wqo-infix-decidable} $\Rightarrow$
    \cref{wqo-prefix-decidable}. We just consider the transduction $f$
    that maps every word $w$ to $\# w$ where $\# $ is a fresh symbol. Then, for
    any language $L \in \mathcal C$, $L$ is \kl{well-quasi-ordered} by
    \kl{prefix} if and only if $f(L)$ is \kl{well-quasi-ordered} by \kl{infix}.
	
    \cref{wqo-prefix-decidable} $\Rightarrow$
    \cref{emptiness-decidable}. 
	We 
	consider the transduction $R \defined \Sigma^* \times \set{a, b}^*$. Then 
	for any language $L \in \mathcal C$,
    the image of $L$ through $R$ is \kl{well-quasi-ordered}
    by \kl{prefix} if and only if $L$ is empty.
\end{proofof}

The class $\mathcal{C}_\text{aut}$ of \kl{regular languages} and the class
$\mathcal{C}_{\text{cfg}}$ of context-free languages are examples of
\kl{effective amalgamative classes}, hence the following corollary.

\begin{corollary}
    \label{aut-cfg-infix:cor}
    Let $\mathcal{C} \in \set{ \mathcal{C}_\text{aut}, \mathcal{C}_{\text{cfg}}}$.
    It is decidable whether a language in $\mathcal{C}$ is \kl{well-quasi-ordered}
    by the \kl{infix relation}.
    Furthermore, whenever it is \kl{well-quasi-ordered} by the \kl{infix relation},
    it is a \kl{bounded language}.
\end{corollary}

Let us conclude by noting that it is unsurprisingly not possible to decide
whether a decidable language is \kl{well-quasi-ordered} by the \kl{prefix
relation}. This is a very easy result whose sole purpose is to contrast with
the decidability result of \cref{aut-cfg-infix:cor}.

\begin{remark}
    The following problem is undecidable: given a language $L$
    decided by a Turing machine, answer whether 
    $L$ is \kl{well-quasi-ordered} for the \kl{prefix relation}.
\end{remark}
\begin{proof}
    We reduce the halting problem on the empty string $\varepsilon$.
    Let $M$ be a Turing Machine, we write the languages $L$ of finite runs
    of $M$ starting on the empty string,
    that we surround by special markers. This language is decidable,
    and 
    is
    \kl{well-quasi-ordered} if and only if it is finite
    if and only if $M$ terminates on $\varepsilon$.
\end{proof}
\section{Conclusion}
\label{conclusion:sec}

We provided concretes statements that justify why the \kl{subword relation}
is used when defining \kl{well-quasi-orders} on finite words. Even if
\kl{prefix}, \kl{suffix} or \kl{infix} relations are meaningful, they are
\kl{well-quasi-ordered} if and only if they behave similarly to disjoint copies
of $\Nat$ or $\Nat^2$. However, our approach suffers some limitations 
and opens the road to natural continuation of this line of work.

\subparagraph{Towards infinite alphabets} In this paper, we restricted our
attention to \emph{finite} alphabets, having in mind the application to
\kl{regular languages}. However, the conclusions of
\cref{bounded-language:thm}, \cref{small-ordinal-invariants:thm}, and
\cref{prefixes:thm} could be conjecture to hold in the case of infinite
alphabets (themselves equipped with a \kl{well-quasi-ordering}). This would
require new techniques, as the finiteness of the alphabet is crucial to all of
our positive results.

\subparagraph{Monoid equations}  It could be interesting to understand which
monoids $M$ recognize languages that are \kl{well-quasi-ordered} by the
\kl{infix}, \kl{prefix} or \kl{suffix} relations. This research direction is
connected to finding which classes of graphs of \emph{bounded clique-width} are
\kl{well-quasi-ordered} with respect to the \emph{induced subgraph relation},
as shown in \cite{DRT10}, and recently revisited by one of the authors in
\cite{L24:arxiv:v2}.

\subparagraph{Complexity} We have chosen to disregard complexity considerations
when proving decidability results, as we do not believe that a fined grained
complexity analysis would be particularly enlightening. However, we conjecture
that for regular languages, the complexity should be polynomial in the size of
the minimal automaton recognizing the language.

\subparagraph{Lexicographic orderings} There is another natural ordering on
words, the \emph{lexicographic ordering}, which does not fit well in our
current framework because it is always of \kl{ordinal width} $1$. However, the
order-type of the lexicographic ordering over \kl{regular languages} has
already been investigated in the context of infinite words \cite{CACOPU18}, and
it would be interesting to see if one can extend these results to decide
whether such an ordering is \kl{well-founded} for languages recognized by
\kl{amalgamation systems}.

\bibliographystyle{plainurl}

\end{document}